\documentclass[aip,graphicx]{revtex4-1}

\usepackage{amsmath}
\usepackage{graphicx}
\usepackage{color}

\begin{document}

\title{The dynamics of an externally driven nanoscale beam that is under high tension and immersed in a viscous fluid}

\newcommand{\BU}{Department of Mechanical Engineering, Division of Materials Science and Engineering, and the Photonics Center, Boston University, Boston, Massachusetts 02215, USA}

\author{J. Barbish}
\affiliation{Department of Mechanical Engineering, Virginia Tech, Blacksburg, Virginia 24061, USA}

\author{C. Ti}
\affiliation{\BU}

\author{K. L. Ekinci}
\affiliation{\BU}

\author{M. R. Paul}
\email{mrp@vt.edu}
\affiliation{Department of Mechanical Engineering, Virginia Tech, Blacksburg, Virginia 24061, USA}

\date{\today}

\begin{abstract}
We explore the dynamics of a nanoscale doubly-clamped beam that is under high tension, immersed in a viscous fluid, and driven externally by a spatially varying drive force. We develop a theoretical description that is valid for all possible values of tension, includes the motion of the higher modes of the beam, and accounts for a harmonic force that is applied over a limited spatial region of the beam near its ends. We compare our theoretical predictions with experimental measurements for a nanoscale beam that is driven electrothermally and immersed in air and water. The theoretical predictions show good agreement with experiment and the validity of a simplified string approximation is demonstrated.

\vspace{2cm}
\noindent \emph{This article may be downloaded for personal use only. Any other use requires prior permission of the author and AIP Publishing. This article appeared in the Journal of Applied Physics and may be found at DOI: 10.1063/5.0100462.}
\end{abstract}

\maketitle

\section{Introduction}
\label{section:introduction}
The dynamics of small elastic beams immersed in a viscous fluid are at the heart of many important technologies~\cite{ekinci:2005,arlett:2011}. Typical devices using doubly-clamped beams composed of silicon nitride posses varying degrees of tension and exhibit a fundamental mode of oscillation with a natural frequency in the megahertz range and a spring constant on the order of 1 N/m.  However, when the beam is immersed in a viscous fluid the frequency of the fundamental peak shifts to lower frequency and the quality factor of the oscillation reduces significantly~\cite{sader:1998,paul:2004,paul:2006}. These reductions in performance become significantly larger when the viscosity and density of the fluid is appreciable as in the case of water. These reductions become even more significant when the dimensions of the beam are uniformly decreased from the microscale down to the nanoscale~\cite{paul:2004,paul:2006}.  These issues have led to interesting solutions such as the use of the higher modes of oscillation to increase the frequency of the measurement~\cite{maali:2005,vaneysden:2007,stachiv:2014}, tailoring of the beam geometry to improve performance~\cite{villa:2009}, using paddle shaped nanoscale cantilevers to drastically reduce stiffness~\cite{paul:2004,paul:2006,arlett:2007}, and placing the fluid of interest inside of the oscillating cantilever rather than immersing the cantilever in the fluid to significantly increase the quality factor~\cite{burg:2007}.

From a broad point of view, device performance for many applications improves if the frequency representing the peak of the amplitude spectrum can be increased. In essence, with an increase in frequency the energy stored by the fluid and solid system increases more than energy lost by dissipation due to the viscous fluid per oscillation~\cite{rosenhead:1963,sader:1998,paul:2006}. The end result is an oscillator with a higher quality. One accessible way to increase the frequency of oscillation, without changing the beam geometry or composition, is to include a tension force~\cite{stachiv:2014}. For example, a doubly clamped beam with tension can be the result of the fabrication process or an intrinsic property of the material~\cite{ari:2021,ti:2021}. Including a tension force in addition to rigidity has several favorable outcomes which can be drawn from the following observations: (i)~the natural frequencies of all of the modes will increase, (ii)~the relative natural frequencies are closer together in the frequency domain, (iii)~the relative stiffness of the different modes are closer together, and (iv)~the quality factor of the oscillations will increase. (ii)-(iii) are important since they reduce the range of measurements that are required in an experiment. (ii)~reduces the frequency range that must be resolved in a experiment. (iii) reduces the required range of beam displacements that must be resolved in an experiment. Lastly, (iv)~is useful since higher quality factors are desirable experimentally due to the more defined peak that results in a measurement of the amplitude as a function of frequency.

The relative magnitude of the peaks in the amplitude spectrum, for the different modes of oscillation, depend upon how the beam is driven and also on the interactions with the surrounding fluid. Small beams can be driven by Brownian motion alone or the beams can also be driven externally. There is significant interest in the stochastic motion of small elastic objects in a viscous fluid~\cite{sader:1998,paul:2004}. However, due to the increased stiffness of the higher modes, the magnitude of the amplitude of oscillations for the higher modes, when driven thermally, are very small.  The theory we describe here could be extended to this case but we do not explore this here. Typically, when a beam is driven externally the magnitude of the deflections are well above the Brownian fluctuations of the beam and the dynamics are treated deterministically. In this paper, we treat this case and focus our attention upon small beams that are driven externally where the external drive force can be tailored to yield favorable magnitudes of oscillation in an experiment. In particular, we consider a doubly-clamped beam that is driven near its attachment points by a spatially varying drive force.

One approach used in the literature to achieve a spatially varying harmonic driving force is electrothermal actuation~\cite{bargatin:2007,ari:2021,ti:2021}. The essential idea is to heat the two ends of the beam using an electric current through a typically U-shaped gold wire by Joule heating. The localized heating causes a temperature gradient that results in a stress gradient due to the differential amount of thermal expansion that occurs in the solid material composing the beam. This generates a bending moment which yields the flexural deflections. By driving the current harmonically in time, it is possible to drive the beam at a chosen frequency and to then sweep over a wide range of frequencies of interest. This approach has been demonstrated to be very effective in driving the flexural oscillations of small beams at frequencies of over 200 MHz~\cite{bargatin:2007}.

In this paper, we quantitatively explore these ideas for a wide range of conditions where we pay particular attention to the role of tension on the beam dynamics and to the influence of the external driving. We explore a doubly-clamped beam that is immersed in a viscous fluid and driven externally by a spatially varying drive force. We develop a theoretical description that is valid for all values of tension in the beam which is bounded by an Euler-Bernoulli beam in the absence of tension and by a string description where tension dominates contributions from elasticity. We compare our theoretical predictions with experimental measurements for an electrothermally driven beam under high tension.

\section{Theory}
\label{section:theory}
\begin{figure}[h]
\begin{center}
\includegraphics[width=3.25in]{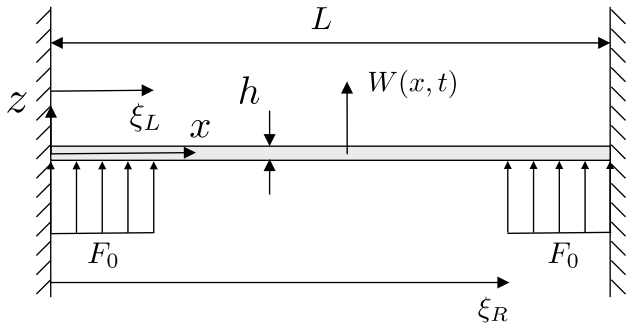}
\end{center}
\caption{A schematic of the doubly-clamped beam with a spatially varying driving force. The beam has a length $L$, width $b$ (into the page), and thickness $h$ with axial coordinate $x$, transverse coordinate $z$, and the remaining coordinate $y$ is into the page. The harmonic driving force has a constant magnitude $F_0$ and is applied near the left and right ends of the beam over the spatial regions specified by the constants $\xi_L$ and $\xi_R$.}
\label{fig:beam}
\end{figure}

We are interested in the dynamics of a microscale elastic beam with tension that is immersed in a viscous fluid and driven externally by a spatially varying harmonic drive force. A schematic is shown in Fig.~\ref{fig:beam} that represents the general configuration we consider. The doubly-clamped beam has length $L$, width $b$, and thickness $h$.  The axial distance along the beam is the $x$ direction, transverse to the beam is the $z$ direction, and the $y$ direction (not shown) is into the page.  The harmonic driving force has a constant magnitude $F_0$ and is applied over the spatial regions specified by the constants $\xi_L$ and $\xi_R$. The flexural displacement of the beam at position $x$ and time $t$ is given by $W(x,t)$.  
%%%%%%%%%%%%%%%%%%%%%%%%%%%%%%%%%%%%%%%%%%%%%%%%%%%%%%%%%%%%%%%%%%%
%%%%%%%%%%%%%%%%%%%%%%%%%%%%%%%%%%%%%%%%%%%%%%%%%%%%%%%%%%%%%%%%%%%

In our study, we will use the beam geometry and material properties given in Table~\ref{table:geom}.  Although specifying a particular beam geometry is not necessary for the development of the theory, it allows us to provide dimensional diagnostics regarding this beam over a wide range of conditions. This beam is also typical of a nanoscale beam fabricated from silicon nitride that has been used in development of new technologies~\cite{ari:2021,ti:2021}. We will quantify the dynamics of this beam as a function of the tension in the beam and for a varying spatial extent of the external forcing. Lastly, this particular beam geometry is precisely the one used in our experimental investigation that is described in Section~\ref{section:experiment}.  By introducing the geometry and material properties now we will tailor our discussion toward the comparison between theory and experiment. In our theoretical development we will pay close attention to the influence of tension, $\xi_L$, $\xi_R$, and $F_0$ on the dynamics of the beam.     
\setlength{\tabcolsep}{12pt}
\begin{table}[h!]
\begin{center}
\begin{tabular}{ c c c c c} 
 $L$ & $b$ & $h$ & $\rho_s$ & $E$ \\
 ($\mu$m) & ($\mu$m) & ($\mu$m) & (kg/m$^3$) & (GPa)  \\ \hline \hline 
 40 & 0.90 & 0.10 & 2960 & 300 \\ 
\end{tabular}
\end{center}
\caption{Geometry and material properties. The doubly-clamped beam has length $L$, width $b$, thickness $h$, density $\rho_s$, and Young's modulus $E$. We will consider this beam immersed in air or water. For the fluid density $\rho_f$ and dynamic viscosity $\mu_f$ we use: for air $\rho_f = 1.23$ kg/m$^3$, $\mu_f = 1.79 \times 10^{-5}~\frac{\text{kg}}{\text{m} \, \text{s}}$; for water $\rho_f = 997.8$ kg/m$^3$, $\mu_f = 9.53 \times 10^{-4}~\frac{\text{kg}}{\text{m} \, \text{s}}$.}
\label{table:geom}
\end{table}

For a long and thin beam, $L \gg b \gg h$, that is under tension, the flexural deflections are described by 
\begin{equation}
E I \frac{\partial^4 W(x,t)}{\partial x^4} - F_T \frac{\partial^2 W(x,t)}{\partial x^2} + \mu \frac{\partial^2 W(x,t)}{\partial t^2} = F_f(x,t) + F_d(x,t)
\label{eq:eom}
\end{equation}
where $E$ is the Young's modulus, $I \!=\!b h^3/12$ is the area moment of inertia, $F_T$ is the tension force, $\mu \!=\! \rho_s b h$ is the mass per unit length, $F_f(x,t)$ is the fluid force per unit length acting on the beam, and $F_d(x,t)$ is the spatially varying driving force per unit length.  In the absence of tension, $F_T\!=\!0$, and Eq.~(\ref{eq:eom}) reduces to an Euler-Bernoulli beam immersed in a fluid. The doubly-clamped beam satisfies fixed boundary conditions such that $W(0,t) \!=\! W(L,t) \!=\! \frac{d W(0,t)}{dx} \!=\!  \frac{d W(L,t)}{dx} \!=\! 0$.  It will be convenient to introduce a nondimensional axial coordinate $x^* \!=\! x/L$ while leaving time as dimensional to yield 
\begin{equation}
\frac{E I}{L^4} \frac{\partial^4 W(x^*,t)}{\partial x^{*4}} - \frac{F_T}{L^2} \frac{\partial^2 W(x^*,t)}{\partial x^{*2}} + \mu \frac{\partial^2 W(x^*,t)}{\partial t^2} = F_f(x^*,t) + F_d(x^*,t).
\label{eq:beam-equation}
\end{equation}
For the boundary conditions this yields $W(0,t)\!=\!W(1,t)\!=\!\frac{\partial W(0,t)}{\partial x^*}\!=\!\frac{\partial W(1,t)}{\partial x^*}\!=\!0$. In what follows we will assume that $x$ is nondimensional and drop the $*$ notation to simplify the notation. 

\subsection{The natural frequencies and mode shapes of a beam with tension}
The analysis of a beam with tension has been described in detail elsewhere~\cite{bargatin:2007,stachiv:2014} and we provide only the essential details in support of our discussion. The natural frequencies and mode shapes are found for the case of no damping and no driving (setting $F_f \!=\! F_d \!=\! 0$ in Eq.~(\ref{eq:beam-equation})) and assuming that for each mode we can express the solution as $W_n(x,t) \!=\! Y_n(x) e^{i \omega_n t}$ where $n$ is the mode number, $Y_n(x)$ is the $n$th mode shape, and $\omega_n$ is the $n$th natural frequency. Substituting this solution into Eq.~(\ref{eq:beam-equation}) for these conditions yields
\begin{equation}
\frac{E I}{L^4} \frac{d^4 Y_n(x)}{d x^4} - \frac{F_T}{L^2} \frac{d^2 Y_n(x)}{d x^2} - \mu \omega_n^2 Y_n(x) = 0
\label{eq:beam-ode}
\end{equation}
which has the solution~\cite{bokaian:1990}
\begin{equation}
Y_n(x) = c_{1,n} \sinh(M_n x) + c_{2,n} \cosh(M_n x) + c_{3,n} \sin(N_n x) + c_{4,n} \cos(N_n x) 
\label{eq:Yn}
\end{equation}
where the mode shapes $Y_n(x)$ are orthogonal, $M_n \! = \! \left( U + \sqrt{U^2 + \Omega_n^2} \right)^{1/2}$, and $N_n \!=\! \left( -U + \sqrt{U^2 + \Omega_n^2} \right)^{1/2}$. The nondimensional tension parameter $U$ is 
\begin{equation}
U = \frac{F_T}{2 E I/L^2}
\label{eq:U}
\end{equation}
which represents a ratio of the tension force to an elastic force scale. The parameter $U$ is very useful in determining the impact of the tension on the beam. For $U\!=\!0$ the Euler-Bernoulli beam result is recovered, for increasing $U$ a beam with tension is described, and for $U \rightarrow \infty$ (or equivalently $E \rightarrow 0$) a string description is recovered. The nondimensional natural frequency for mode $n$ is
\begin{equation}
\Omega_n = \frac{\omega_n}{\alpha/L^2}
\label{eq:Omega}
\end{equation}
where $\alpha \!=\! (EI/\mu)^{1/2}$. Inserting Eq.~(\ref{eq:Yn}) into Eq.~(\ref{eq:beam-ode}) and rearranging yields the characteristic equation~\cite{bokaian:1990} 
\begin{multline}
\Omega_n + U \sinh\left[ \left( U + \sqrt{U^2 + \Omega_n^2} \right)^{1/2} \right] \sin\left[ \left( -U + \sqrt{U^2 + \Omega_n^2}\right)^{1/2} \right] \\ - \Omega_n \cosh \left[ \left( U + \sqrt{U^2 + \Omega_n^2}\right)^{1/2} \right] \cos \left[ \left(-U + \sqrt{U^2 + \Omega_n^2} \right)^{1/2} \right]  = 0
\label{eq:characteristic-equation}
\end{multline}
where
\begin{eqnarray}
c_{1,n} &=& 1, \\ 
c_{2,n} &=& \frac{M_n \sin (N_n) - N_n \sinh (M_n)}{N_n \left[ \cosh(M_n) - \cos(N_n)\right]}, \\
c_{3,n} &=&  - \frac{M_n}{N_n},   \\ 
c_{4,n} &=& - c_{2,n}.
\end{eqnarray}
For a given value of the tension parameter $U$, the roots of Eq.~(\ref{eq:characteristic-equation}) yield the nondimensional natural frequencies $\Omega_n$. It will be useful to define the normalized mode shapes as  $\phi_n(x) \!=\! \frac{1}{\sqrt{\tilde{n}_n}} Y_n(x)$ where $\tilde{n}_n$ is a normalization constant whose value is given by
\begin{equation}
\tilde{n}_n = \int_{0}^{1} Y_n(x) Y_n(x) dx.
\end{equation}
The orthonormal mode shapes $\phi_n(x)$ therefore satisfy
\begin{equation}
\int_{0}^{1} \phi_n(x) \phi_m(x) dx = 
\begin{cases}
1 ~~~~~~~~~~n = m \\ 
0 ~~~~~~~~~~n \ne m  
\end{cases}
\end{equation}
The mode shapes $\phi_n$ of an Euler-Bernoulli beam without tension are recovered for $U\!=\!0$ and the mode shapes for a beam with tension are quantified using a finite value of $U$.

The limit of the tension parameter $U \! \rightarrow \! \infty$ indicates the dominance of tension over elasticity which can also be represented as $E \! \rightarrow \! 0$. Neglecting the elastic contribution by setting $E\!=\!0$ in Eq.~(\ref{eq:beam-equation}) yields the string equation 
\begin{equation}
- \frac{F_T}{L^2} \frac{\partial^2 W(x,t)}{\partial x^2} + \mu \frac{\partial^2 W(x,t)}{\partial t^2} = F_f(x,t) + F_d(x,t)
\label{eq:string2}
\end{equation}
with the no-displacement boundary conditions $W(0,t) \!=\! W(1,t) \!=\! 0$ where the slope of the string at the boundaries may be non-zero. The mode shapes and natural frequencies of the string can be found by analyzing Eq.~(\ref{eq:string2}) in the absence of the fluid interaction force and the driving force to yield $\phi_n(x) \!=\! - \sqrt{2} \sin(n \pi x)$ and $\omega_n \!=\! \frac{\pi c}{L} n$ where $c \!=\! \sqrt{F_T/\mu}$ is the wave speed for the string and the minus sign is to match the convention used when describing the modes of a beam in Eq.~(\ref{eq:Yn}). This yields $\omega_n/\omega_1 \!=\! n$ indicating that the relative natural frequencies of the string increase linearly with $n$.
%%%%%%%%%%%%%%%%%%%%%%%%%%%%%%%%%%%%%%%%%%%%%%%%%%%%%%%%%%%%%%%%%%%
\begin{figure}[h]
\begin{center}
\includegraphics[width=2.1in]{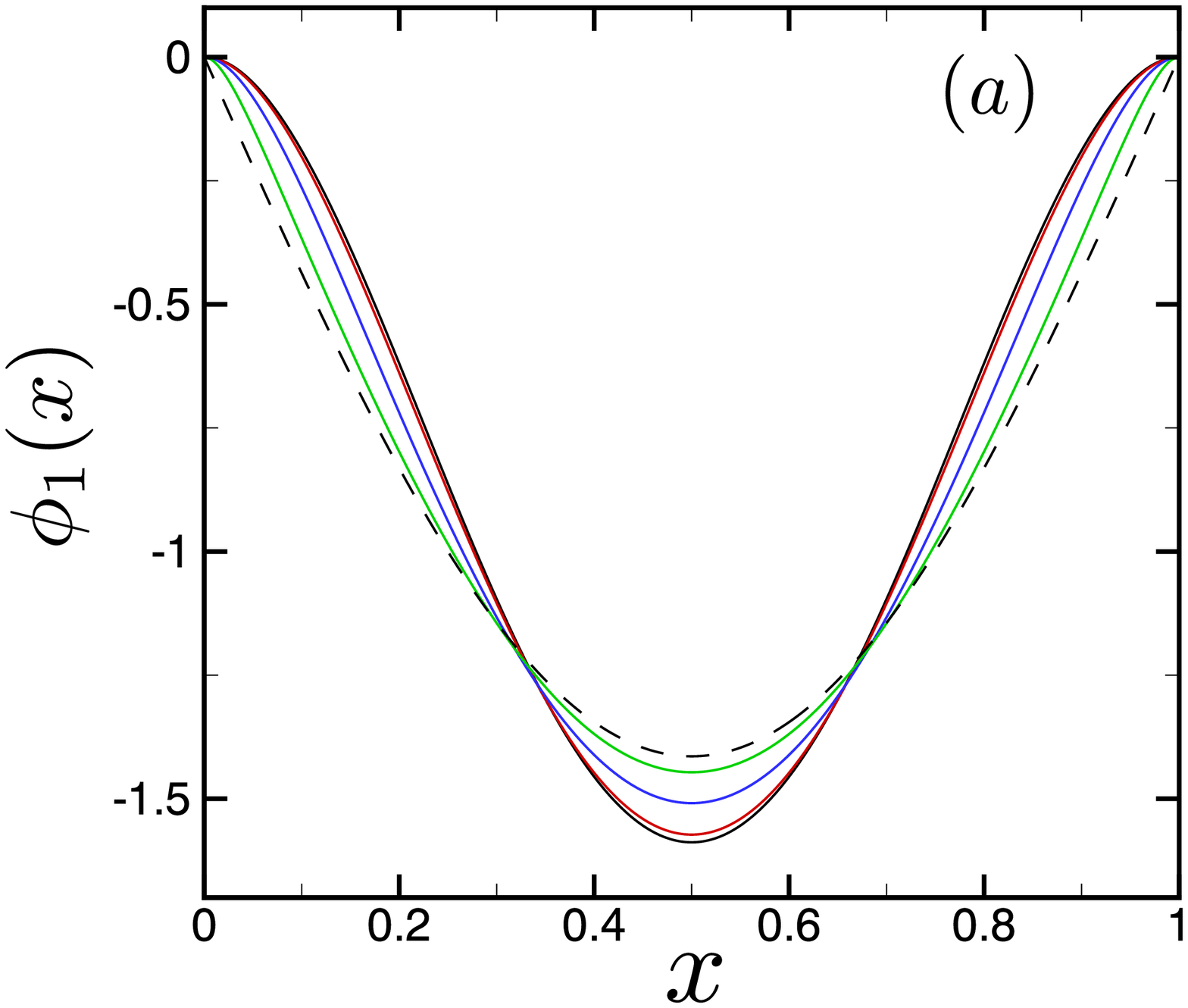}
\includegraphics[width=2.1in]{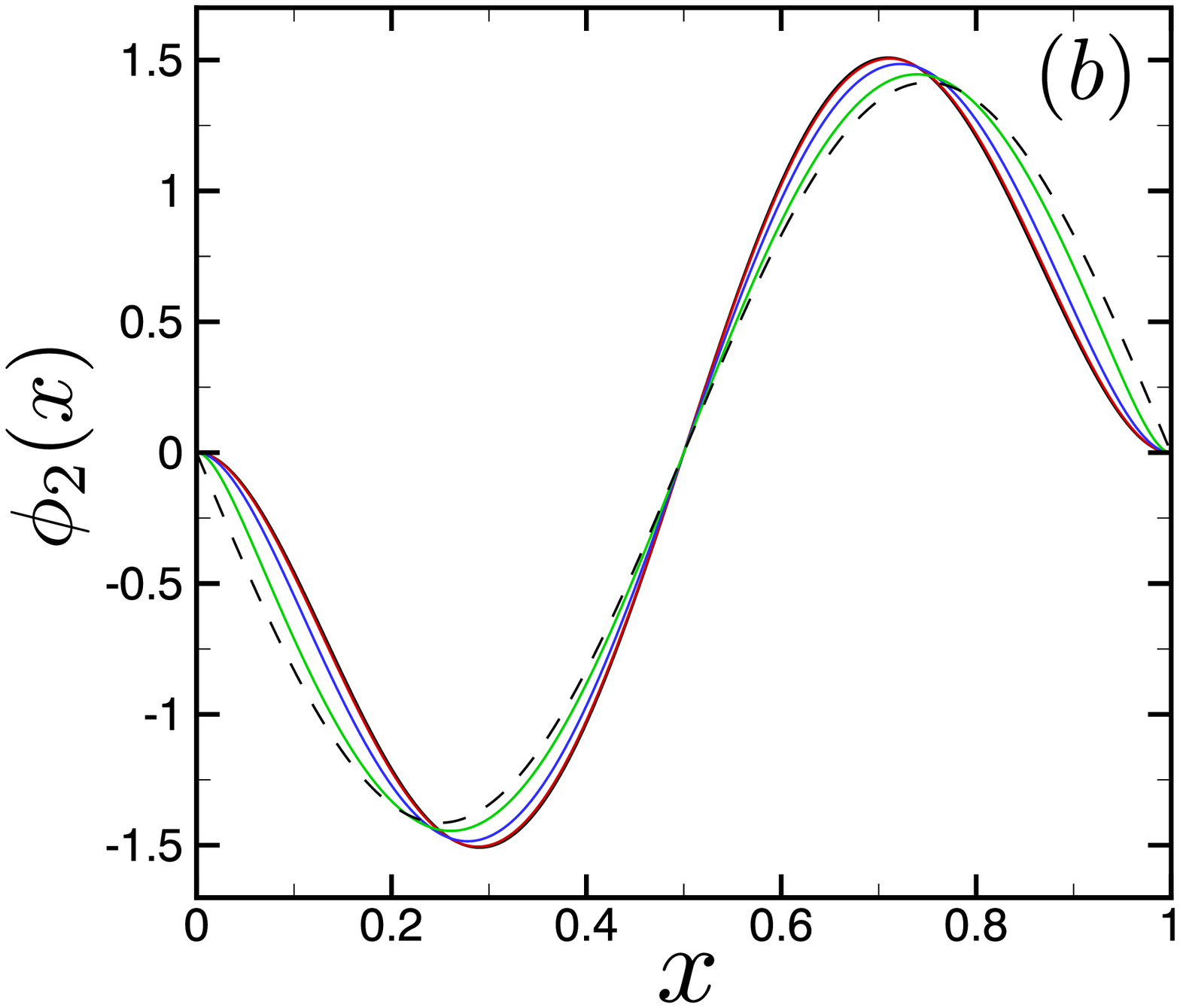}  
\includegraphics[width=2.1in]{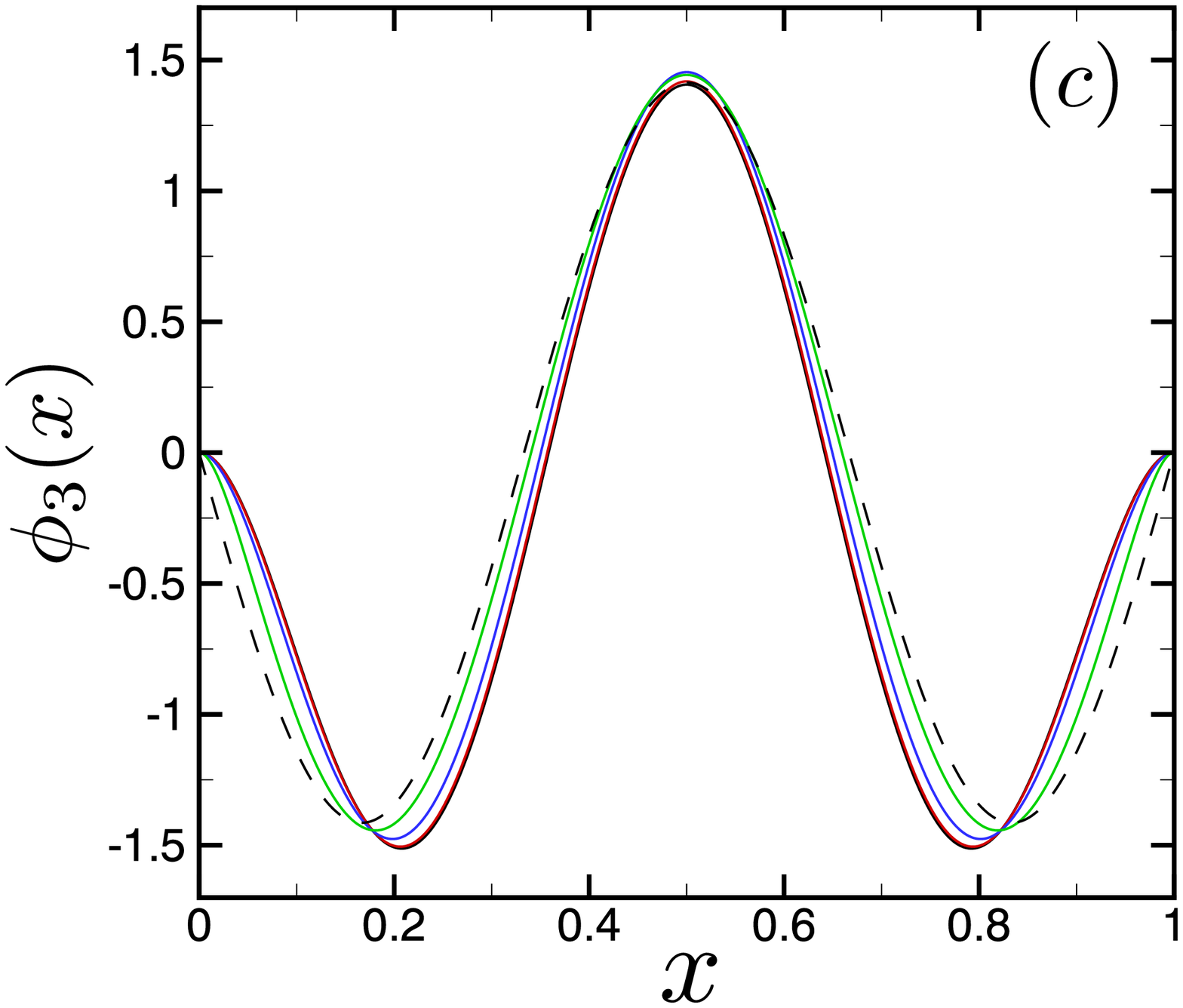} \\
\vspace{0.5cm}
\includegraphics[width=2.1in]{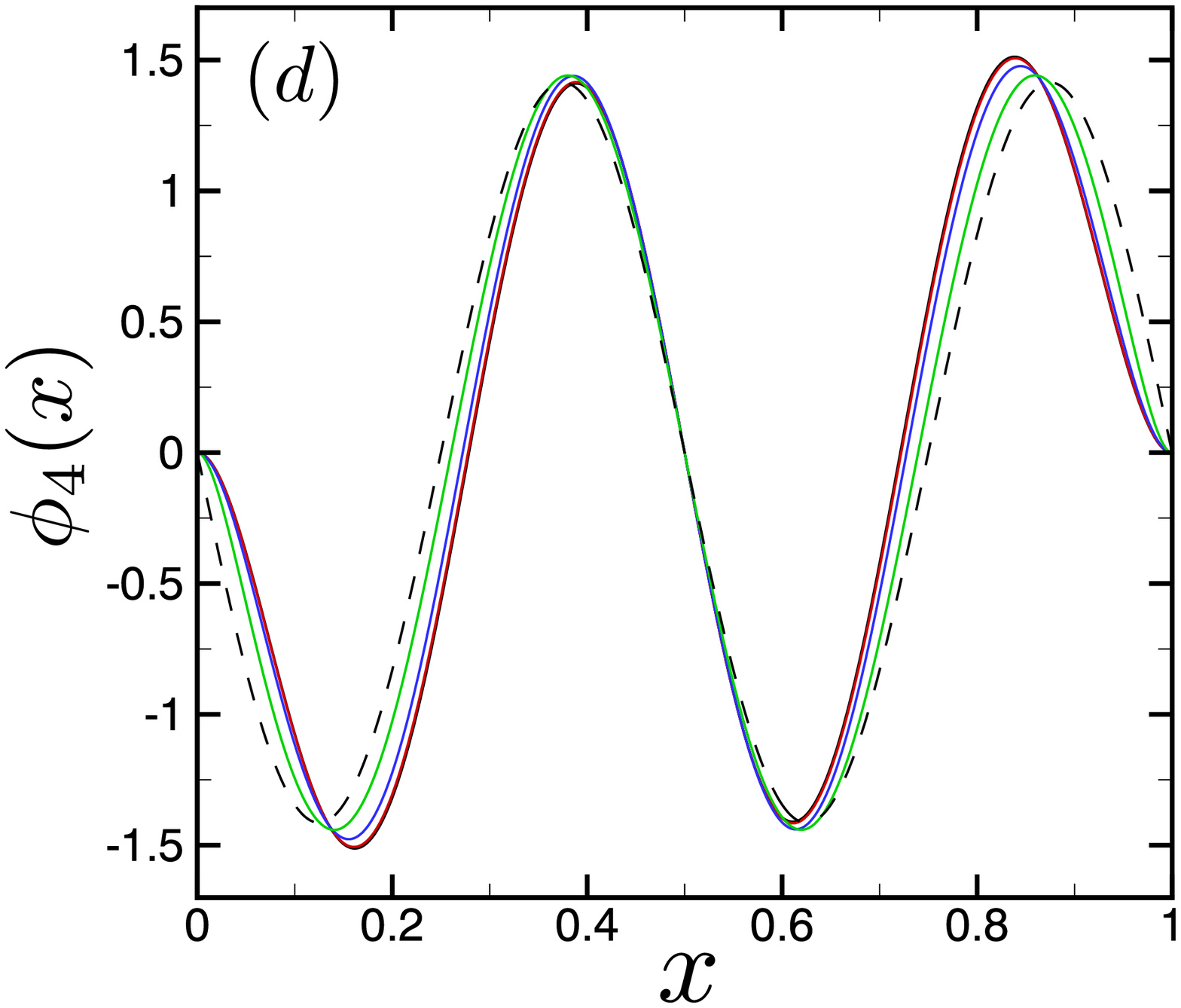} 
\includegraphics[width=2.1in]{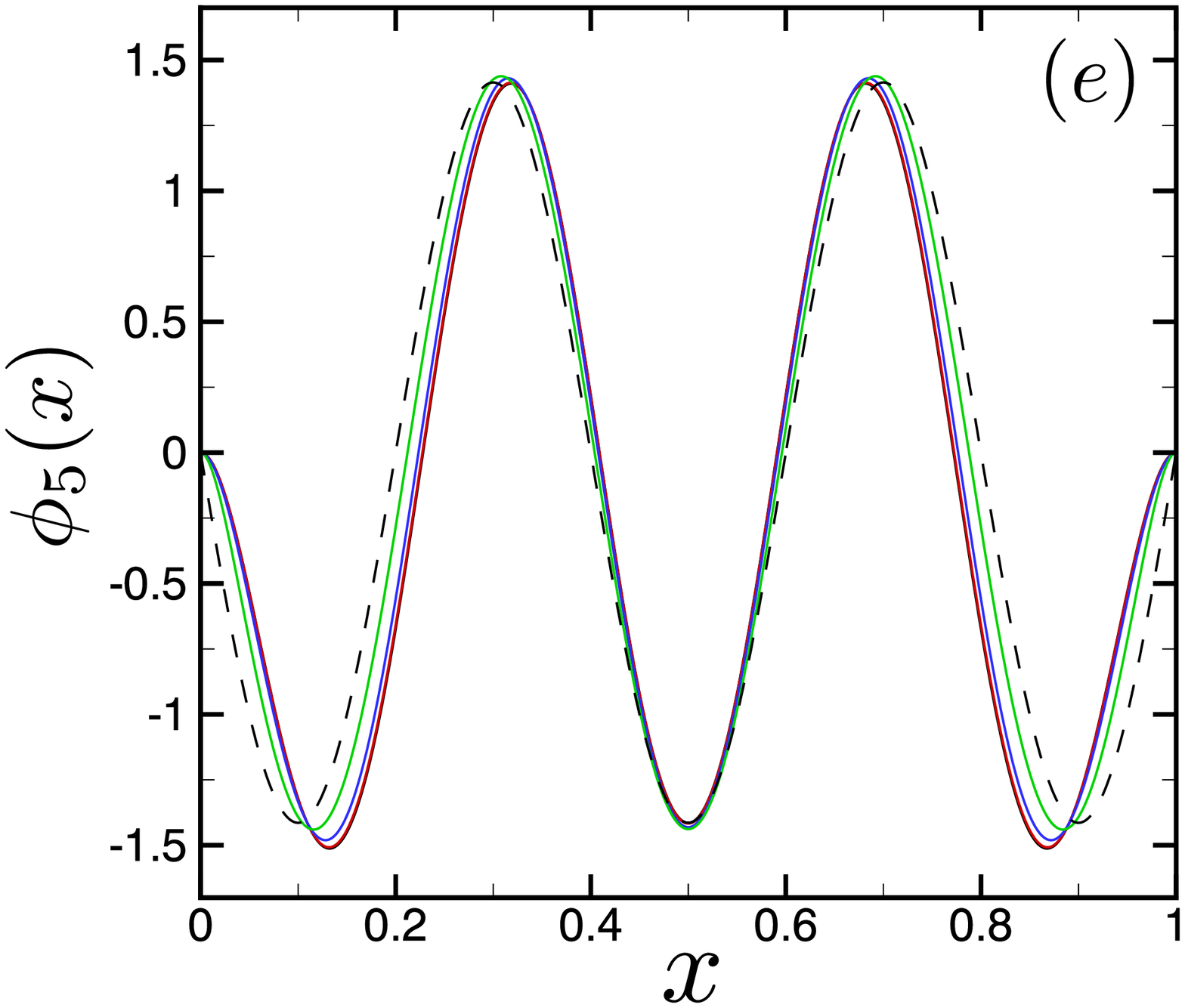}
\end{center}
\caption{The first five orthonormalized mode shapes $\phi_n(x)$ as a function of the amount of tension in the beam. (a)-(e) show modes 1 through 5, respectively. Each panel has 5 curves: the Euler-Bernoulli beam $U=0$ (black, solid), $U\!=\!10$ (red), $U\!=\!100$ (blue), $U\!=\!1000$ (green) and the limit of a string (dashed). In all panels the solid lines approach the dashed line as the tension is increased.}
\label{fig:modes-all}
\end{figure}
%%%%%%%%%%%%%%%%%%%%%%%%%%%%%%%%%%%%%%%%%%%%%%%%%%%%%%%%%%%%%%%%%%%

The first five mode shapes $\phi_n(x)$ are shown in Fig.~\ref{fig:modes-all} as a function of $U$. Curves are included for an Euler-Bernoulli beam without tension $U=0$, as the solid black line, beams with increasing tension where $U\!=\!10$ (red), 100 (blue), 1000 (green), and a string shown as the dashed line. In all panels, the curves approach the string result as the tension is increased. It is clear that the tension affects the mode shapes.  This is very evident at the boundaries where the beam must approach the walls horizontally while the string does not.  This is of particular interest because it is near these boundaries, at $x\!=\!0$ and $x\!=\!L$, where the spatially varying external drive force will be applied. Therefore it is anticipated that the relative magnitudes of the peaks in the amplitude spectrum will depend upon the details of the mode shapes and their variation with tension since this will affect how the external driving is coupled with the beam motion.

\begin{figure}[h]
\begin{center}
\includegraphics[width=3in]{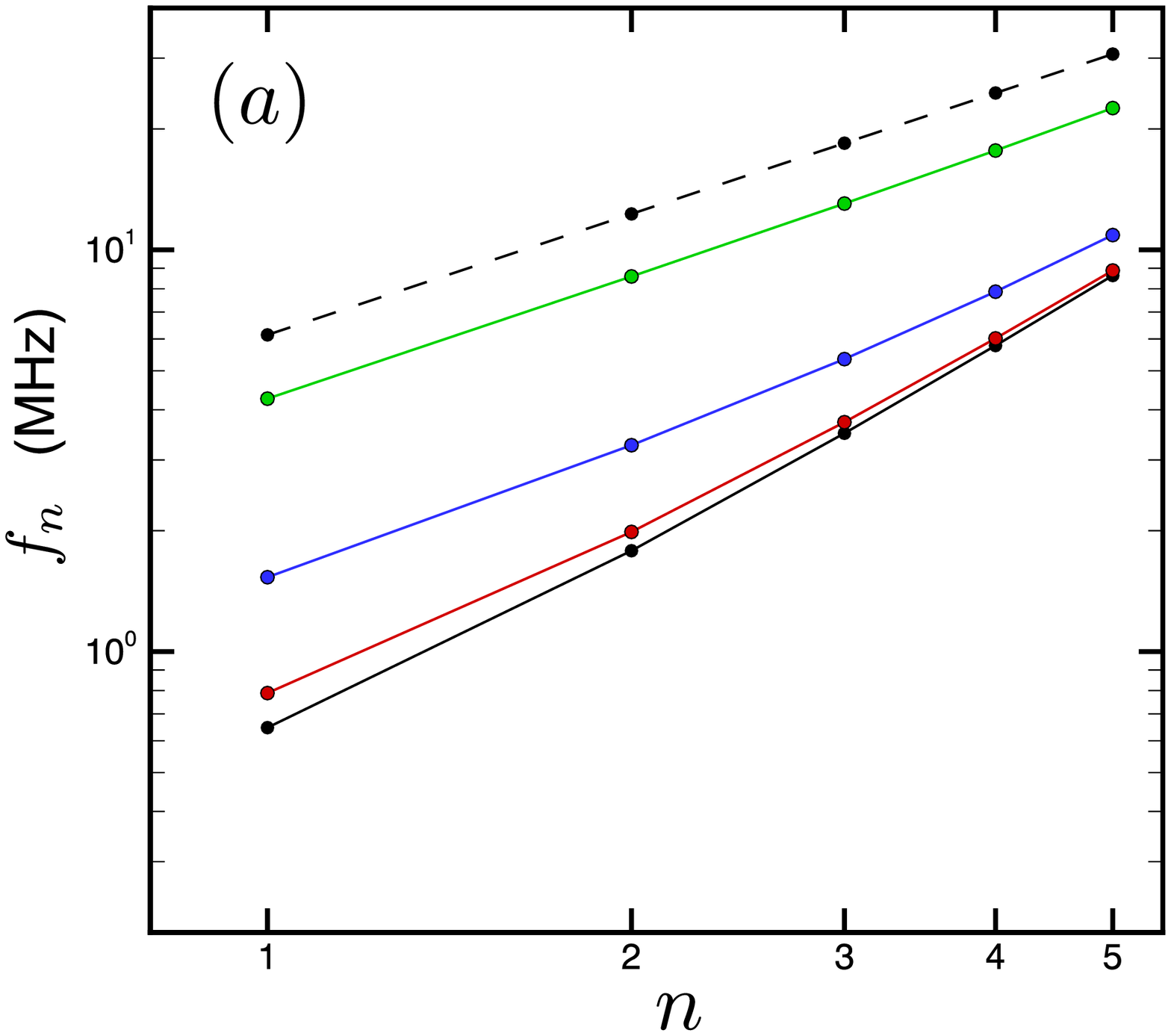}
\includegraphics[width=3in]{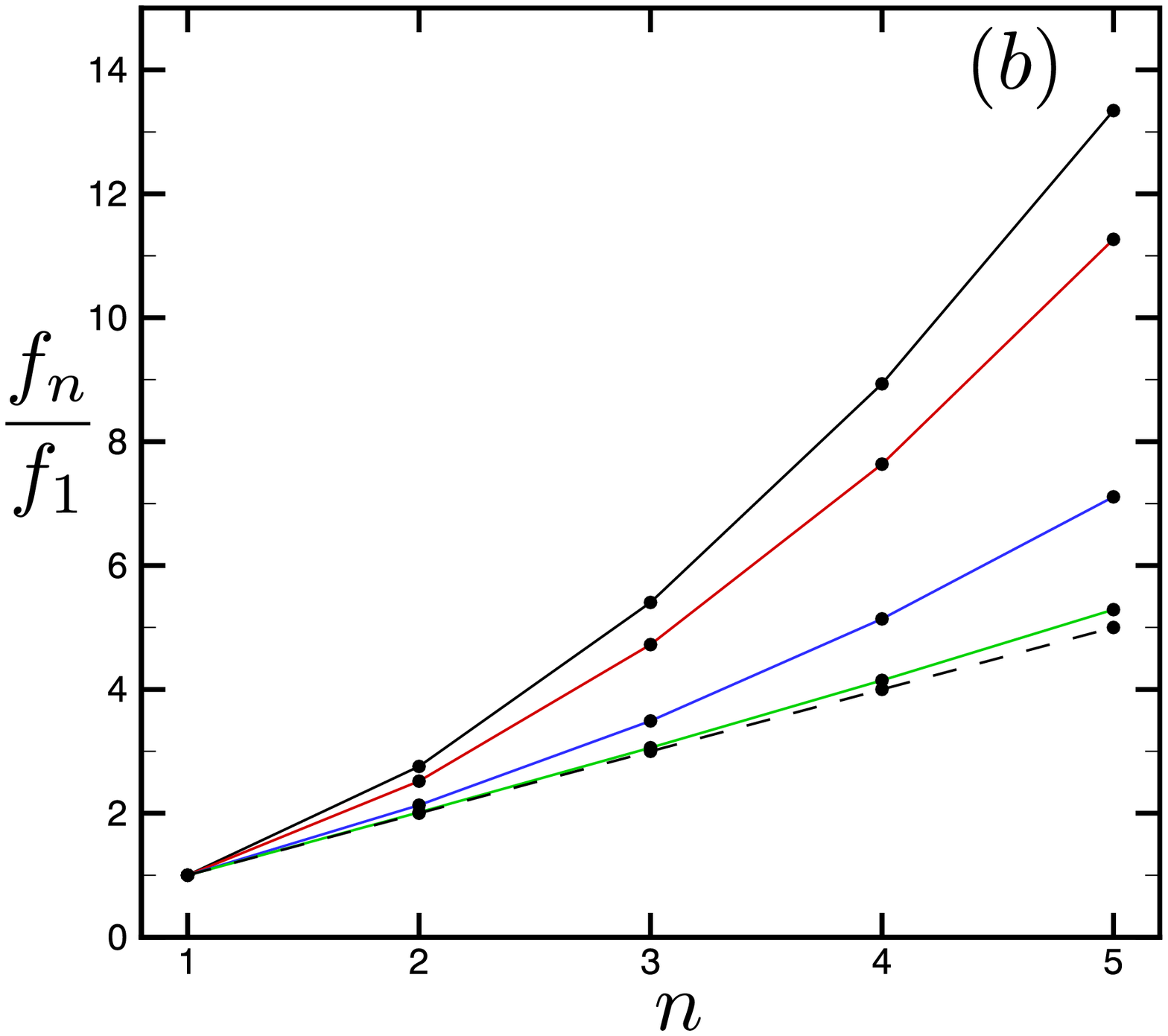}  
\end{center}
\caption{(a) A log-log plot of the variation of the first five natural frequencies $f_n$ of the beam with the mode number $n$ as a function of the tension parameter $U$ where $U\!=\!0$ (black, solid), $U\!=\!10$ (red), $U\!=\!100$ (blue), $U\!=\! 1000$ (green) and a string (dashed). (b) The same data plotted as $f_n/f_1$ versus $n$. In both panels the curves approach the string result in order with increasing $U$.}
\label{fig:freq}
\end{figure}

The variation of the natural frequencies of the beam with tension is shown in Fig.~\ref{fig:freq}. Curves are included for $U=0,10,100,1000$ and for a string (dashed line). The frequency of oscillation increases significantly with increasing tension and with increasing mode number. In particular, the fifth mode of the beam under very high tension has a natural frequency of over 30 MHz. As the tension is increased, the variation of frequencies approach the linear trend of the string.  This is shown more clearly in Fig.~\ref{fig:freq}(b) which plots the variation of the normalized frequency $f_n/f_1$ as a function of $n$. This illustrates how the relative separation of the frequencies decreases toward the linear behavior as $U$ is increased.
%%%%%%%%%%%%%%%%%%%%%%%%%%%%%%%%%%%%%%%%%%%%%%%%%%%%%%%%%%%%%%%%%%%
%%%%%%%%%%%%%%%%%%%%%%%%%%%%%%%%%%%%%%%%%%%%%%%%%%%%%%%%%%%%%%%%%%%

Once the mode shapes are known, it is straightforward to determine the effective mass $m_n$ and effective spring constant $k_n$ of the modes in the usual manner\cite{paul:2006} by ensuring the kinetic energy and potential energy of the entire beam are captured by the lumped mode when measured at some position $x_0$. This yields $m_n \!=\! \alpha_n m$ where $\alpha_n \!=\! \phi_n(x_0)^{-2}$. The effective spring constant of mode $n$, when the measurement of the displacement is made at $x_0$, is then $k_n \!=\! m_n \omega_n^2$.  When these ideas are applied to the string this yields $k_n \!=\! \frac{\pi^2  F_T}{L \phi_n(x_0)^2} n^2$ which shows that $k_n$ increases linearly with $F_T$ at fixed $n$ and quadratically with $n$ at fixed $F_T$ which can also be expressed as $k_n/k_1 \!=\! n^2$.

\begin{figure}[h]
\begin{center}
\includegraphics[width=3in]{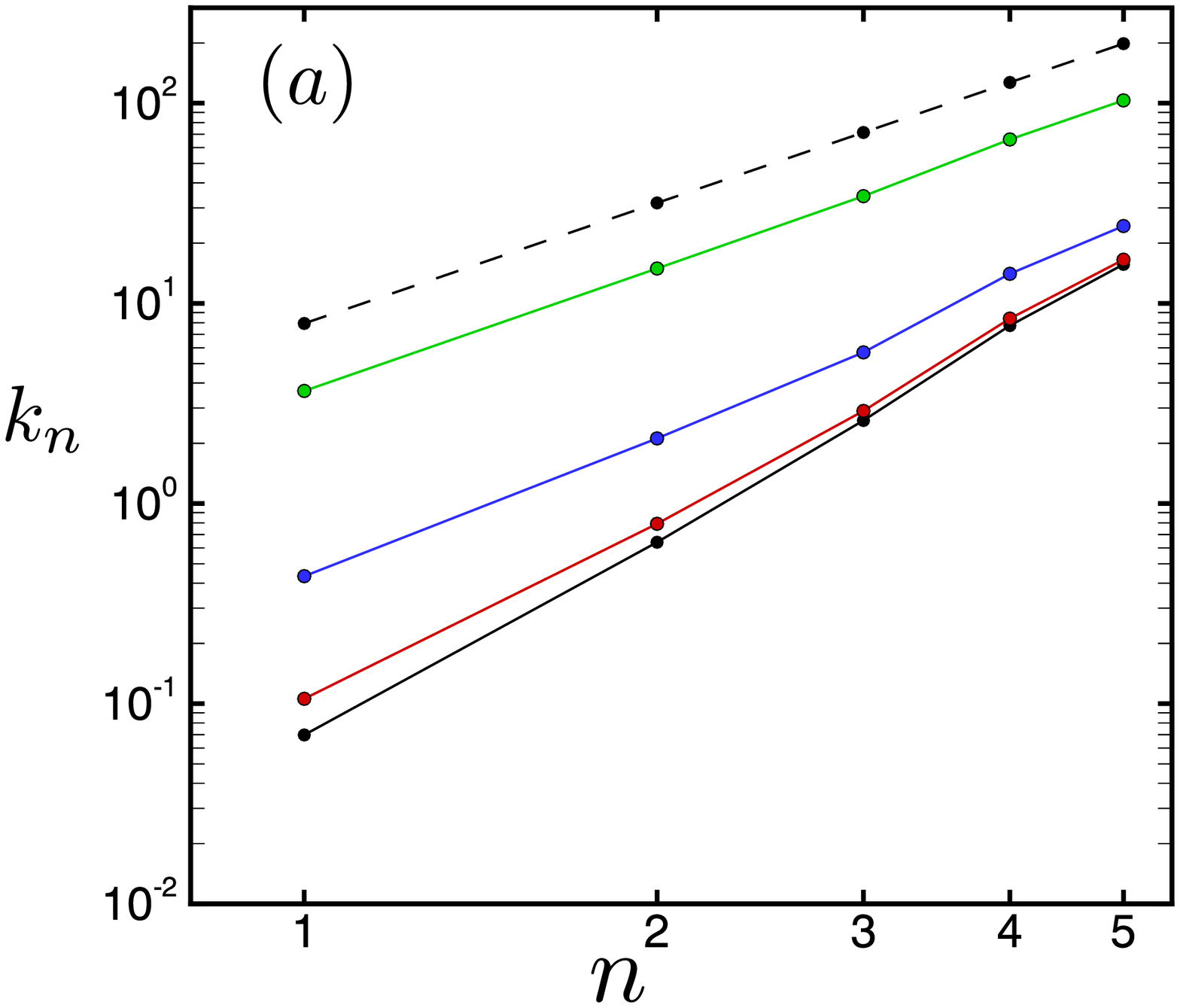}
\includegraphics[width=3in]{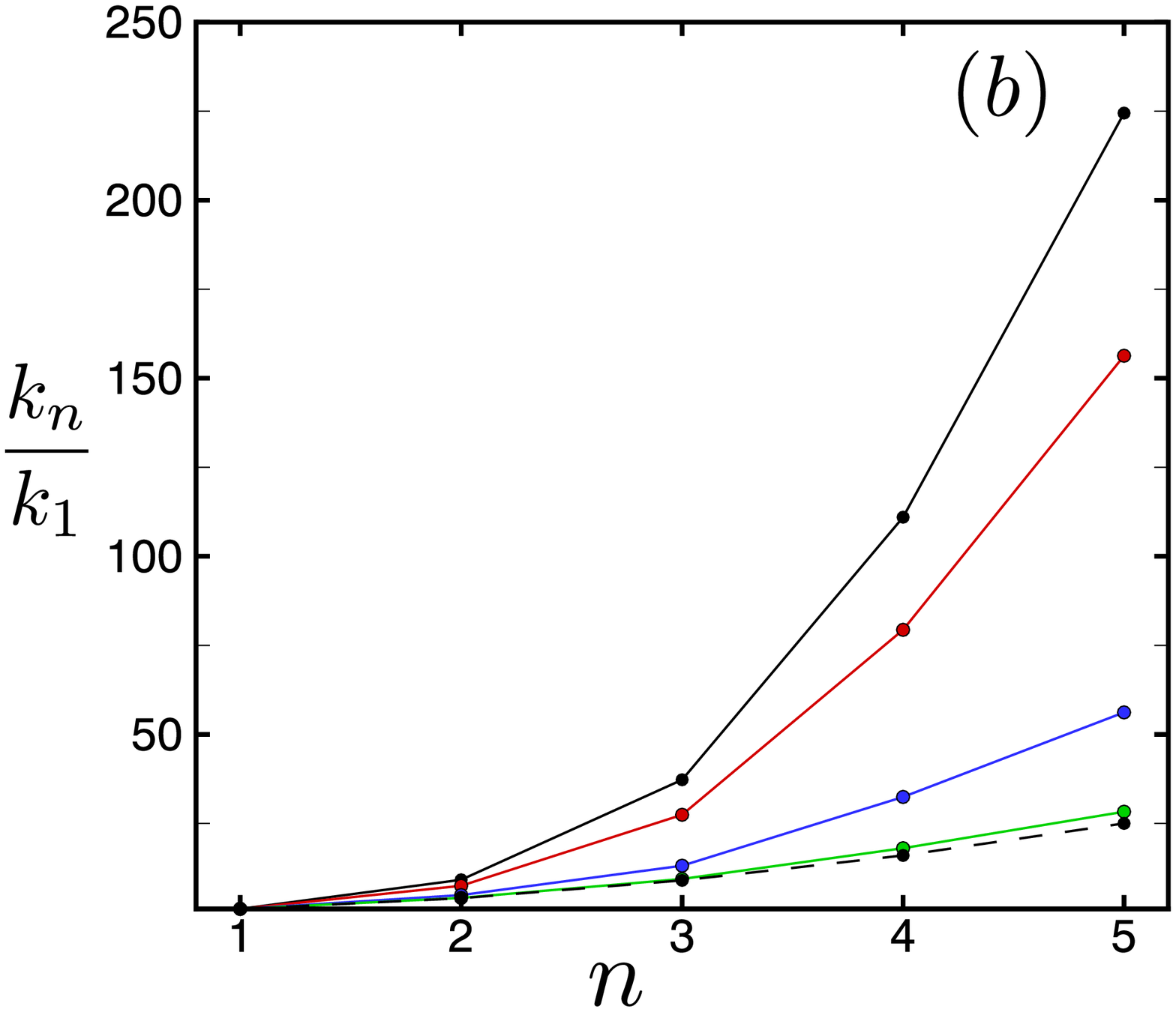} 
\end{center}
\caption{(a)~A log-log plot of the first five spring constants $k_n$ with varying tension where $U\!=\!0$ (black), $U\!=\!10$ (red), $U\!=\!100$ (blue), $U\!=\!1000$ (green), and a string (dashed) in units of N/m. For the odd modes $k_n$ is measured at $x_0\!=\!1/2$, for the second mode $x_0\!=\!1/4$, and for the fourth mode $x_0 \!=\! 1/8$. (b)~The same data plotted as $k_n/k_1$ versus $n$. In both panels, the results approach the string in order with increasing values of $U$.}
\label{fig:k-all}
\end{figure}

The variation of $k_n$ with $n$ is shown in Fig.~\ref{fig:k-all}(a) as a function of $U$. In order to present the results of all of the modes on a single plot it must be understood that the measurement is taken for each mode at the location $x_0$ of an antinode where that mode shape $\phi_n(x)$ is maximum. Specifically, $x_0\!=\!1/2$ for the odd modes, $x_0\!=\!1/4$ for mode 2, and $x_0\!=\!1/8$ for mode 4. The dramatic increase in the stiffness of the beam with increasing tension is evident by noting that at high tension $k_5 \approx 200$ N/m which is 25 times stiffer than the fundamental mode. The relative magnitude of $k_n$ with respect to $k_1$ is represented in Fig.~\ref{fig:k-all}(b). As the tension is increased, the separation of the spring constants with respect to $k_1$ decrease until reaching the quadratic trend of the string. The relative amplitude of the oscillation of a particular mode with respect to the amplitude of the fundamental mode is given by $k_1/k_n$. When viewed in this light, this indicates that the relative magnitude of the oscillations of the higher modes, with respect to the fundamental mode, decrease with increasing tension. 
%%%%%%%%%%%%%%%%%%%%%%%%%%%%%%%%%%%%%%%%%%%%%%%%%%%%%%%%%%%%%%%%%%%
%%%%%%%%%%%%%%%%%%%%%%%%%%%%%%%%%%%%%%%%%%%%%%%%%%%%%%%%%%%%%%%%%%%

\subsection{The dynamics of an externally driven beam with tension in a viscous fluid}
We now develop the solution for the dynamics of the beam with tension in a viscous fluid that is being driven externally.  It will be convenient to transform Eq.~(\ref{eq:beam-equation}) into frequency space using the Fourier transform pair
\begin{eqnarray}
\hat{W}(x,\omega) &=& \int_{-\infty}^{\infty} W(x,t) e^{i \omega t} dt \\
W(x,t) &=& \frac{1}{2 \pi} \int_{-\infty}^{\infty} \hat{W}(x,\omega) e^{-i \omega t} d \omega
\end{eqnarray}
to yield
\begin{equation}
\frac{E I}{L^4} \frac{\partial^4 \hat{W}(x,\omega)}{\partial x^{4}} - \frac{F_T}{L^2} \frac{\partial^2 \hat{W}(x,\omega)}{\partial x^{2}} - \mu \omega^2 \hat{W}(x,\omega) = \hat{F}_f(x,\omega) + \hat{F}_d(x,\omega).
\label{eq:beam-fourier}
\end{equation}
The force due to the fluid can be expressed as\cite{sader:1998,rosenhead:1963}
\begin{equation}
\hat{F}_f(x,\omega) = \frac{\pi}{4} \rho_f \omega^2 b^2 \Gamma(\omega) \hat{W}(x,\omega)
\label{eq:fluid-force}
\end{equation}
where $\Gamma(\omega)$ is the complex-valued hydrodynamic function for an oscillating blade of width $b$ in a viscous fluid of density $\rho_f$. The hydrodynamic function contains contributions due to the mass loading captured by its real part and due to viscous damping captured by its imaginary part. $\Gamma(\omega)$ can be expressed as $\Gamma(\omega) \!=\! \Omega_c(\omega) \Gamma_c(\omega)$ where $\Gamma_c(\omega)$ is the hydrodynamic function of an oscillating cylinder with diameter $b$ and $\Omega_c(\omega)$ is a complex valued correction factor. Explicit expressions for $\Gamma_c(\omega)$ and $\Omega_c(\omega)$ are given by Sader~\cite{sader:1998}. In Eq.~(\ref{eq:fluid-force}) we have also assumed that $\Gamma(\omega)$ is independent of the mode number $n$ which is expected to be a good approximation for the first several harmonics of the beam. Acoustic radiation and axial flows may cause the dissipation to deviate from the cylinder solution, but these are negligible for lower harmonics \cite{doi:10.1063/5.0037959}. The generalized hydrodynamic function for arbitrary mode number\cite{vaneysden:2006,vaneysden:2007} could be included if desired.

Equation~(\ref{eq:fluid-force}) is valid when the Reynolds number describing the fluid motion,  $\text{Re}$, is small.   For the case of microscale and nanoscale elastic structures in fluid, $\text{Re} \! \ll \! 1$ due to the small amplitudes of the oscillation even though the oscillation frequency is large. The Reynolds number of the fluid motion generated by the motion of mode $n$ of the beam can be expressed as
\begin{equation}
\text{Re}_n = \frac{\rho_f A_n \omega_{n,f} b}{2 \mu_f}
\end{equation}
where $A_n$ is the maximum amplitude of the motion of the $n$th mode, $\omega_{n,f}$ is the frequency of oscillation at which $A_n$ occurs in the fluid, $\mu_f$ is the dynamic viscosity of the fluid, and the length scale has been chosen to be the beam half width $b/2$. For example, in experiment the magnitude of the driving force is typically set to achieve an amplitude of oscillation on the order of $A_1 \approx 1$ nm and when the fluid is air we have the additional simplification $\omega_{n,f} \!\approx\! \omega_n$. If we assume the beam in Table~\ref{table:geom} has high tension and use the string predictions as a guide, this leads to $\text{Re}_1 \!=\! 1.1 \!\times\! 10^{-3}$. Furthermore, the Reynolds number of the higher modes decreases with increasing mode number. Although the quantitative details change when the beam is placed in a more viscous fluid such as water where $\omega_{n,f} \!<\! \omega_n$, the Reynolds number of the beam remains small and it also decreases with increasing mode number.

The quality factor of the oscillations $Q$ can quantified using the hydrodynamic function. Extending the approach used in Ref.\cite{paul:2006} for the fundamental mode to describe the quality of mode $n$ yields
\begin{equation}
    Q_n \approx \frac{\frac{1}{T_0} + \Gamma_r(\omega_{n,f})}{\Gamma_i(\omega_{n,f})}
\end{equation}
where $\Gamma_r$ and $\Gamma_i$ are the real and imaginary parts of the hydrodynamic function, respectively. The frequency dependence of the added mass due to the fluid motion is given by $\Gamma_r (\omega)$ and the frequency dependence of the viscous damping due to the fluid is given by $\omega \Gamma_i(\omega)$. The mass loading parameter, $T_0 \!=\! \frac{\pi \rho_f b}{4 \rho_s h}$, represents the mass of a cylinder of fluid with diameter $b$ to the mass of the beam. The quality $Q_n$ is evaluated at the frequency $\omega_{n,f}$ that yields the maximum amplitude of mode $n$ when driven in fluid.  The quality $Q_n$ increases with increasing values of the frequency $\omega_{n,f}$. Therefore the quality factor will increase with the addition of tension and also with increasing mode number $n$.

\begin{figure}[h]
\begin{center}
\includegraphics[width=3in]{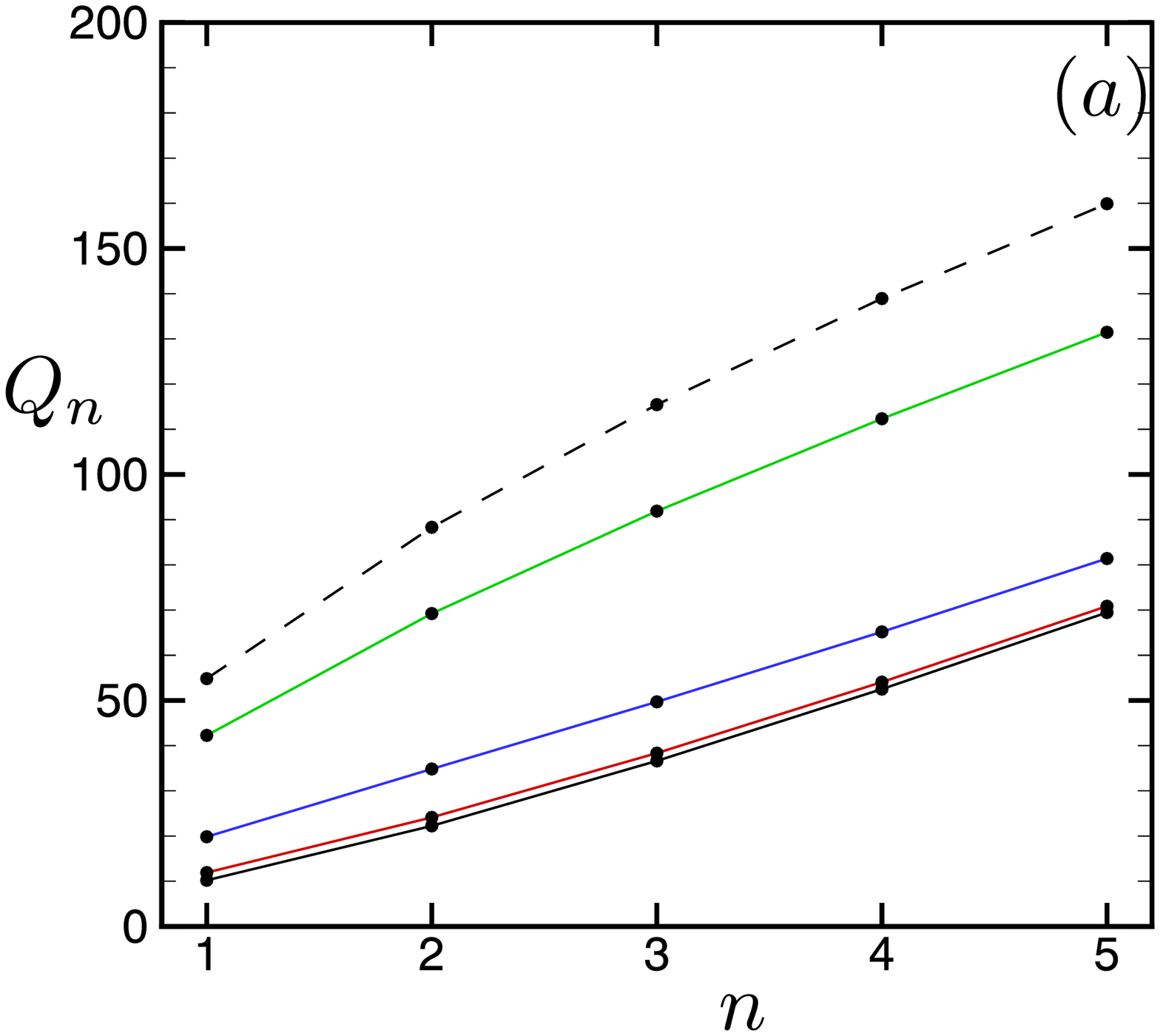}
\includegraphics[width=3in]{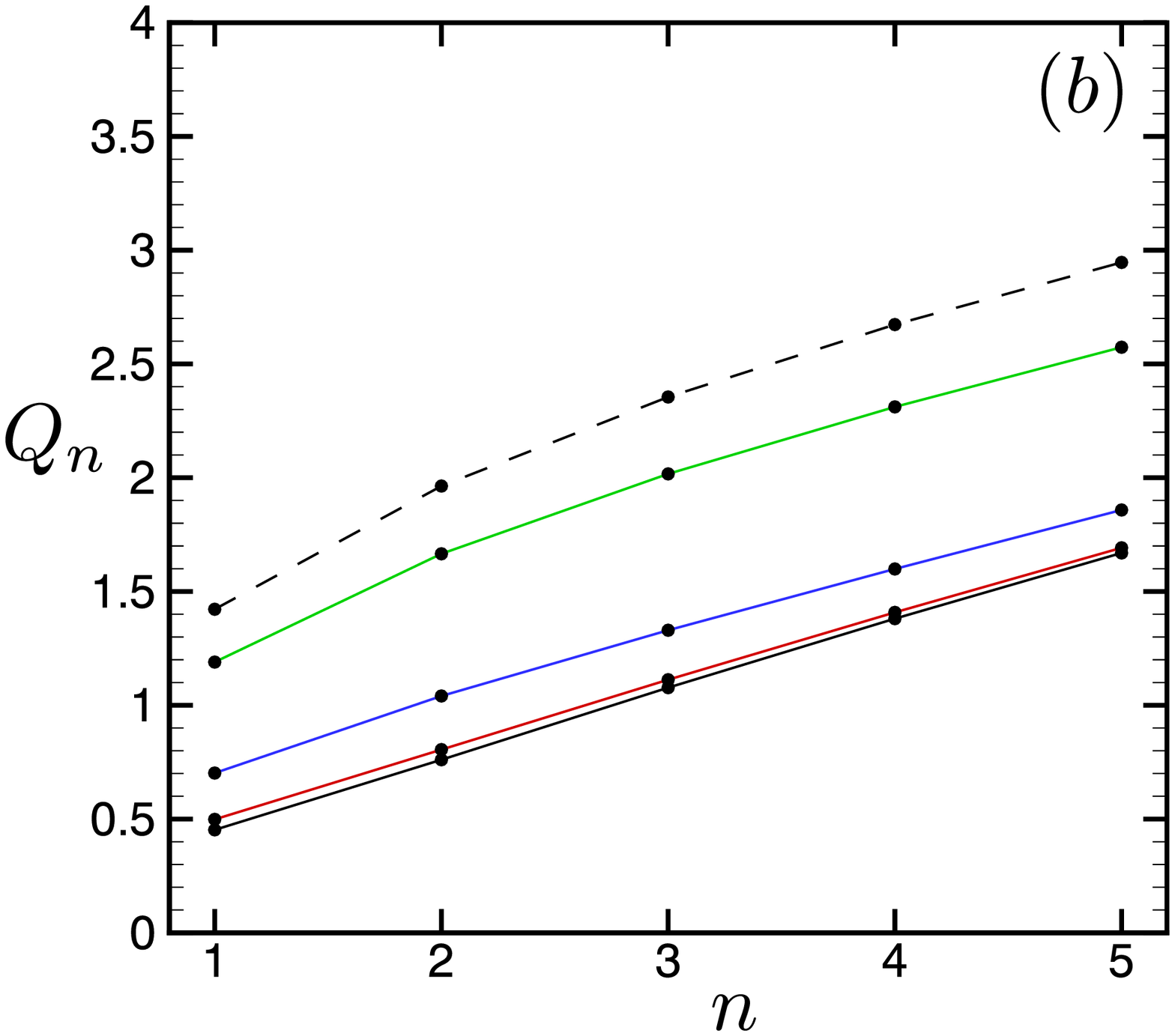}
\end{center}
\caption{The variation of the quality $Q_n$ with mode number $n$ as a function of tension for the beam of Table~\ref{table:geom} immersed in (a)~air and (b)~water. Solid lines are shown for $U\!=\!0$ (black), 10 (red), 100 (blue), 1000 (green) and the dashed line is for a string. The solid lines approach the string result in order as the tension increases.}
\label{fig:q-all}
\end{figure}

The variation of $Q_n$ with tension is shown quantitatively in Fig.~\ref{fig:q-all} for the beam of Table~\ref{table:geom} immersed in (a)~air and in (b)~water.  Moving from left to right in Fig.~\ref{fig:q-all} illustrates the increase in the quality factor as a function of mode number and moving vertically illustrates the increase in the quality factor with increasing tension. Increasing the mode number and the tension significantly increases the quality factor of the oscillations. For example, in Fig.~\ref{fig:q-all}(a), $Q_n$ goes from approximately 10 for the first mode of an Euler-Bernoulli beam to over 150 for the 5th mode of a string. Figure~\ref{fig:q-all}(b) shows the dramatic reduction in quality that occurs when the fluid is water. However, for water it can be seen that $Q_1 \!\lesssim\! 0.5$ for the Euler-Bernoulli beam and that the quality factor increases to nearly $Q_1 \!\approx\! 1.5$ by increasing the tension.
%%%%%%%%%%%%%%%%%%%%%%%%%%%%%%%%%%%%%%%%%%%%%%%%%%%%%%%%%%%%%%%%%%%
%%%%%%%%%%%%%%%%%%%%%%%%%%%%%%%%%%%%%%%%%%%%%%%%%%%%%%%%%%%%%%%%%%%

Following the approach described in Refs.\cite{sader:1998,clark:2010}, we solve Eq.~(\ref{eq:beam-fourier}) using an expansion in terms of the normalized beam modes
\begin{equation}
\hat{W}(x,\omega) = \sum_{n=1}^{\infty} \hat{W}_n(x,\omega) = \sum_{n=1}^{\infty} \hat{f}_n(\omega) \phi_n(x)
\label{eq:expansion}
\end{equation}
where $\hat{W}_n(x,\omega)$ is the complex amplitude of mode $n$ and $\hat{f}_n(\omega)$ describes the frequency dependent amplitude of mode $n$. Inserting Eq.~(\ref{eq:expansion}) into Eq.~(\ref{eq:beam-equation}) and using the orthogonality property of the beam modes yields
\begin{equation}
\hat{f}_n(\omega) \omega_n^2  - \omega^2 \left(1 + T_0  \Gamma_c(\omega) \right) \hat{f}_n(\omega) = \frac{1}{\mu} \int_0^1 \phi_n(x) \hat{F}_d(x,\omega) dx
\end{equation}
which can be solved for $\hat{f}_n(\omega)$. After rearranging, the solution for $\hat{f}_n(\omega)$ can be expressed as 
\begin{equation}
\hat{f}_n(\omega) = \frac{L^4}{E I} \frac{\int_0^1 \hat{F}_d(x,\omega) \phi_n(x) dx} {C_n^4  - B(\omega)^4} 
\label{eq:fn}
\end{equation}
where 
\begin{equation}
B(\omega) = C_1 \left(\frac{\omega}{\omega_1} \right)^{1/2} \left[1 + T_0  \Gamma(\omega) \right]^{1/4}.
\label{eq:B}
\end{equation}
and $C_n = \Omega_n^{1/2}$. Using the final result for $\hat{f}_n(\omega)$ allows us to express the solution for the flexural oscillations as 
\begin{equation}
\hat{W}(x,\omega) = \frac{L^4}{E I} \sum_{n=1}^{\infty} \frac{\int_0^1 \phi_n(x') \hat{F}_d(x',\omega) dx'}{C_n^4  - B(\omega)^4} \phi_n(x).
\end{equation}
The magnitude of the flexural oscillations of the beam measured at position $x_0$ is then $|\hat{W}(x_0,\omega)|$.

A string description can be obtained following a similar approach by starting with Eq.~(\ref{eq:string2}) and using the expression for the fluid force to yield
\begin{equation}
- \frac{F_T}{L^2} \frac{\partial^2 \hat{W}(x,\omega)}{\partial x^2} - \omega^2 \left[ \mu + \frac{\pi}{4} \rho_l b^2 \Gamma(\omega) \right] \hat{W}(x,\omega) = \hat{F}_d(x,\omega).
\end{equation} 
This equation can be solved using an eigenfunction expansion, the mode shapes of the string, and orthogonality of the mode shapes to yield
\begin{equation}
\hat{f}_n(\omega) = \frac{L^2} {\pi^2 F_T} \frac{ \int_0^1 \hat{F}_d(x,\omega) \phi_n(x)  dx} {n^2 - B^2(\omega)}
\label{eq:fnbeam}
\end{equation}
where
\begin{equation}
B(\omega) = \left( \frac{\omega}{\omega_1} \right) \left[ 1 + T_0 \Gamma(\omega) \right]^{1/2}.
\label{eq:Bstring}
\end{equation}
We note that the expression for $B(\omega)$ in Eq.~(\ref{eq:Bstring}) is different for the string when compared to the expression for the beam given by Eq.~(\ref{eq:B}). However, it can be seen that $(B(\omega)/C_1)^2$ of the beam equals $B(\omega)$ of the string. The string displacement is then 
\begin{equation}
\hat{W}(x,\omega) = \frac{L^2} {\pi^2 F_T} \sum_{n=1}^{\infty} \frac{ \int_0^1 \hat{F}_d(x',\omega) \phi_n(x') dx'} {n^2 - B^2(\omega)} \phi_n(x).
\label{eq:whatall-string}
\end{equation}
If the explicit expressions for the string mode shapes are used this can also be expressed as
\begin{equation}
\hat{W}(x,\omega) = \frac{2 L^2} {\pi^2 F_T} \sum_{n=1}^{\infty} \frac{ \int_0^1 \hat{F}_d(x',\omega) \sin(n \pi x')  dx'} {n^2 - B^2(\omega)} \sin(n \pi x).
\end{equation}

\subsection{Modeling a spatially varying drive force}
\label{section:spatially-varying-driving-force}
We next consider a driving force that is constant in magnitude that is applied harmonically at the two edges of the beam  (see Fig.~\ref{fig:beam}). We will include the possibility where the driving force at the ends of the beam can be in-phase or out-of-phase with respect to each other. On the left side of the beam near the wall where $0 \! \le \! x \! \le \! \xi_L$ we have $F_d(x,t) \! = \! a_L \frac{F_0}{L} \sin (\omega_d t)$. Similarly, on the right side of the beam near the wall where $\xi_R \!\le\! x \!\le\! 1$ we have $F_d(x,t) \!=\! a_R \frac{F_0}{L} \sin (\omega_d t )$. Elsewhere on the beam the driving force is not applied. The constants $a_L$ and $a_R$ are $\pm1$ to indicate if the driving at the two ends are in-phase ($a_L\!=\!a_R\!=\!1$) or out-of-phase ($a_L\!=\!1$,  $a_R\!=\!-1$). Due to the symmetry of the mode shapes, the odd modes are even about the middle of the beam ($x\!=\!1/2$), and they are driven with an in-phase drive. The even modes are odd about the middle of the beam and they are driven by an out-of-phase drive. In our analysis, we use an impulse force in time to study the frequency response of the system due to a harmonic drive. In this case, the external drive force is represented as   
\begin{equation}
F_d(x,t) =
\frac{F_0}{L} \begin{cases} 
a_L \delta(t)  ~~~~ 0 \le x \le \xi_L \\ 
0 ~~~~~~~~~~~ \xi_L < x < \xi_R \\ 
a_R \delta(t) ~~~~ \xi_R \le x \le 1\\ 
\end{cases}
\end{equation}
where $\delta(t)$ is the Dirac delta function. In the frequency domain the drive force is then 
\begin{equation}
\hat{F}_d(x,\omega) =
\frac{F_0}{L} \begin{cases} 
a_L ~~~~ 0 \le x \le \xi_L \\ 
0 ~~~~~~ \xi_L < x < \xi_R \\ 
a_R ~~~~ \xi_R \le x \le 1.
\end{cases}
\end{equation}
The coupling of the spatially extended drive force with the beam mode shapes is captured by the integral term in Eq.~(\ref{eq:fn}). If we next assume that the spatial variation of the drive force is of equal spatial extent $\xi^*$, on either side of the beam, we can simplify the notation further by noting  $\xi_L \! = \! \xi^*$ and $\xi_R \!=\! 1 \!-\! \xi^*$. Therefore, $\xi^*$ can vary over the range $0 \!\le\! \xi^* \!\le\! 1/2$ where $\xi^*\!=\!1/2$ corresponds to a driving force that has been applied over the entire length of the beam. If we define the integral term in Eq.~(\ref{eq:fn}) as $\psi_n$, and express the limits of integration using $\xi^*$, we have  
\begin{equation}
\psi_n(\xi^*) = a_L \int_0^{\xi^*} \phi_n(x) dx  + a_R \int_{1-\xi^*}^1 \phi_n(x) dx
\label{eq:psi}
\end{equation}
and
\begin{equation}
\hat{W}(x,\omega) = \frac{F_0 L^3}{E I} \sum_{n=1}^{\infty} \frac{\psi_n(\xi^*) \phi_n(x)}{C_n^4  - B(\omega)^4}
\label{eq:whatfinal-beam}
\end{equation}
where the complex amplitude of mode $n$ is 
\begin{equation}
\hat{W}_n(x,\omega) = \left( \frac{F_0 L^3}{E I} \right) \frac{\psi_n(\xi^*) \phi_n(x)}{C_n^4  - B(\omega)^4}.
\label{eq:whatnfinal-beam}
\end{equation}
The variable $\psi_n(\xi^*)$ quantifies the magnitude of the coupling of the spatial drive force with the individual modes of the beam. The tension in the beam is accounted for by the variation of the mode shape $\phi_n$ with the applied tension.

For the string, using Eq.~(\ref{eq:psi}) yields
\begin{equation}
\hat{W}(x,\omega) = \frac{F_0 L} {\pi^2 F_T} \sum_{n=1}^{\infty} \frac{ \psi_n(\xi^*) \phi_n(x)  } {n^2 - B^2(\omega)}
\end{equation}
where $\psi_n(\xi^*)$ can be evaluated to yield
\begin{equation}
\psi_n(\xi^*) = - \frac{2 \sqrt{2}}{n \pi} \left( 1 - \cos (n \pi \xi^*) \right).
\label{eq:psi-string}
\end{equation}
Equation~(\ref{eq:psi-string}) is valid for even and odd modes of the string where it is assumed that the odd modes have been driven by a symmetric drive and the even modes have been driven by an asymmetric spatial drive. Gathering these results together for the string, we can represent the oscillations in a more convenient form as   
\begin{equation}
\hat{W}(x,\omega) = \frac{4 L F_0} {\pi^3 F_T} \sum_{n=1}^{\infty} \frac{1 - \cos (n \pi \xi^*)} { n \left( n^2 - B^2(\omega) \right)} \sin(n \pi x)
\label{eq:whatfinal-string}
\end{equation}
where the displacement of mode $n$ is 
\begin{equation}
\hat{W}_n(x,\omega) = \left(\frac{4 L F_0} {\pi^3 F_T} \right) \frac{1 - \cos (n \pi \xi^*)} { n \left( n^2 - B^2(\omega) \right)} \sin(n \pi x).
\label{eq:whatnfinal-string}
\end{equation}

\begin{figure}[h]
\begin{center}
\includegraphics[width=2.1in]{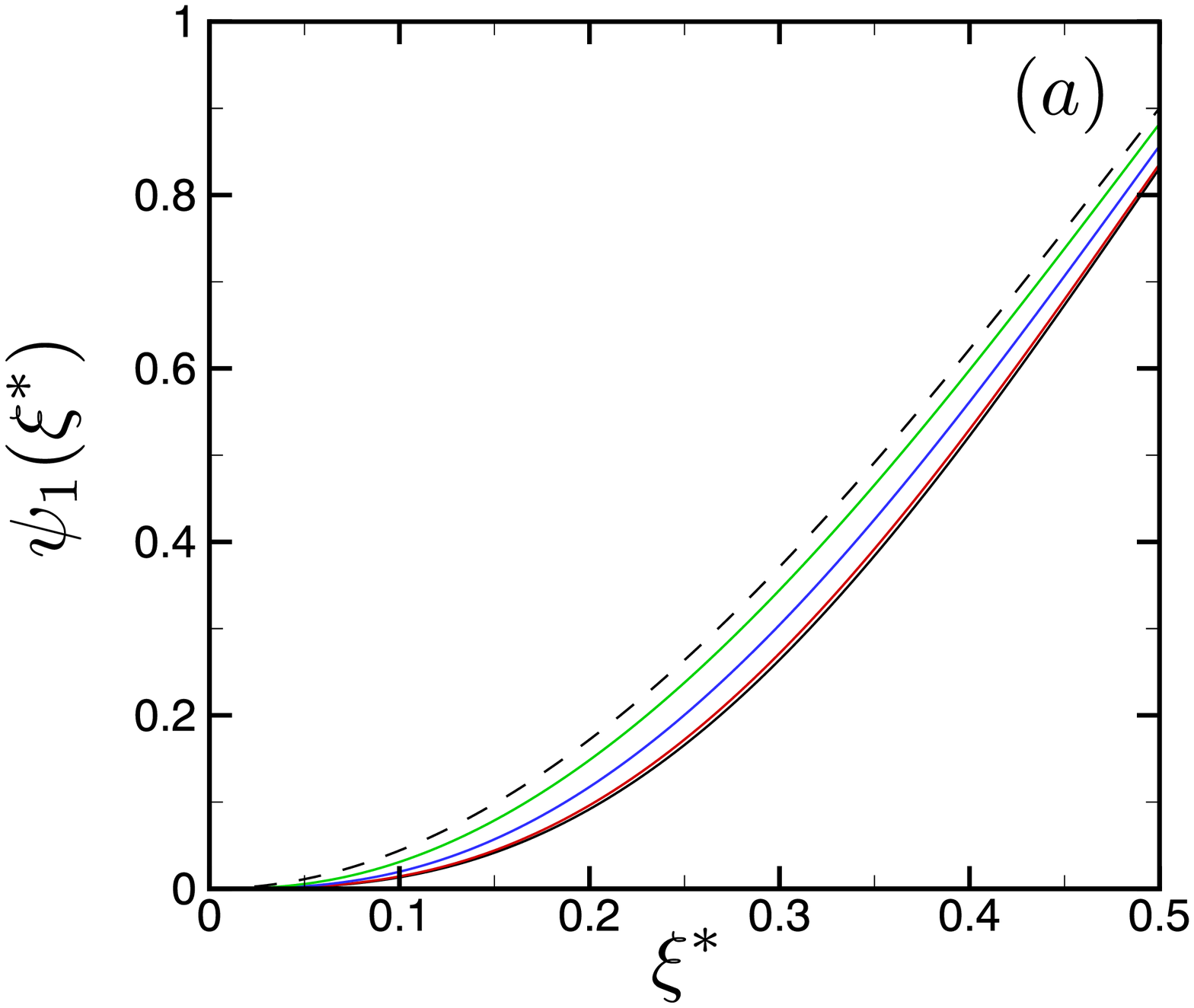}
\includegraphics[width=2.1in]{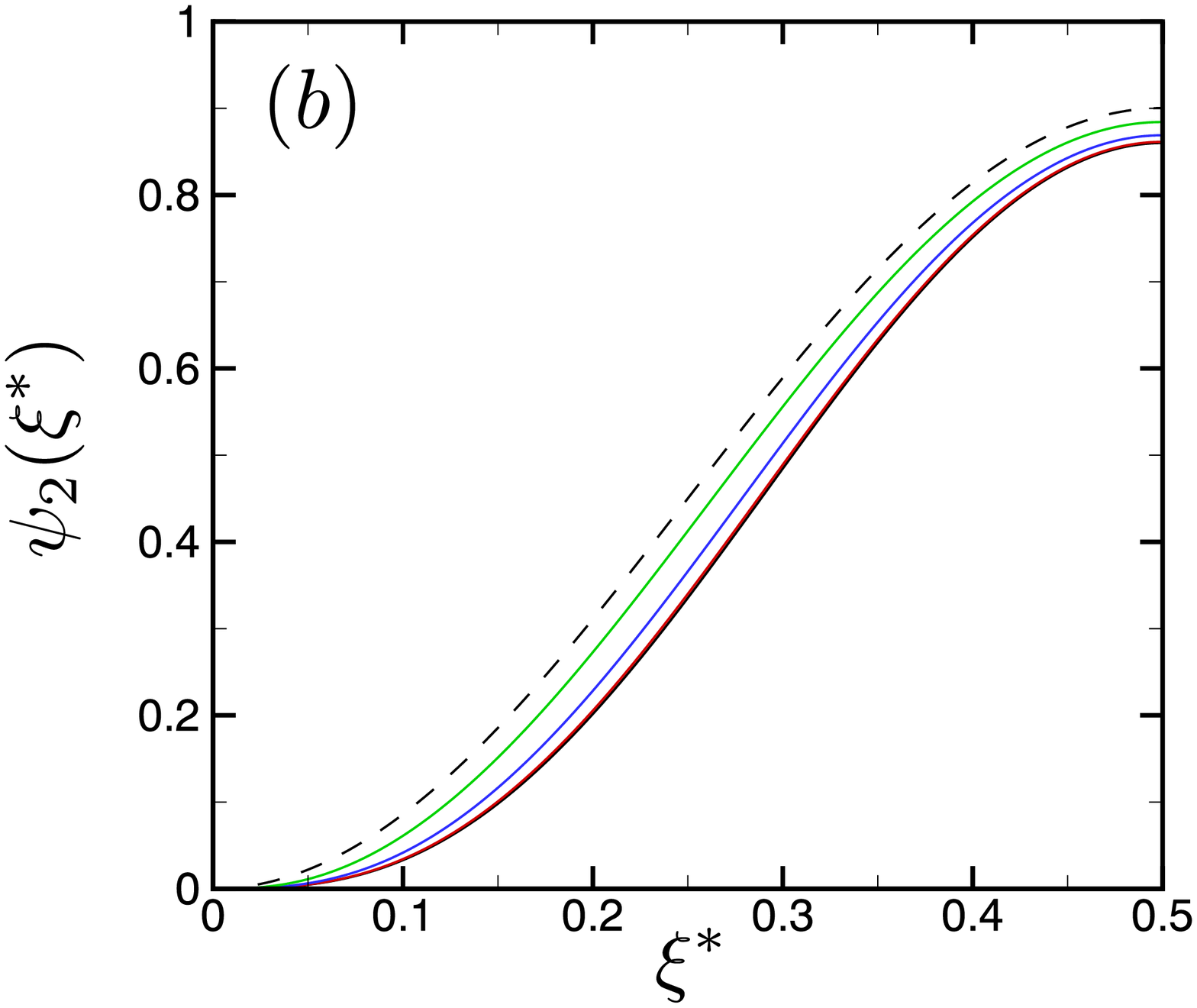}
\includegraphics[width=2.1in]{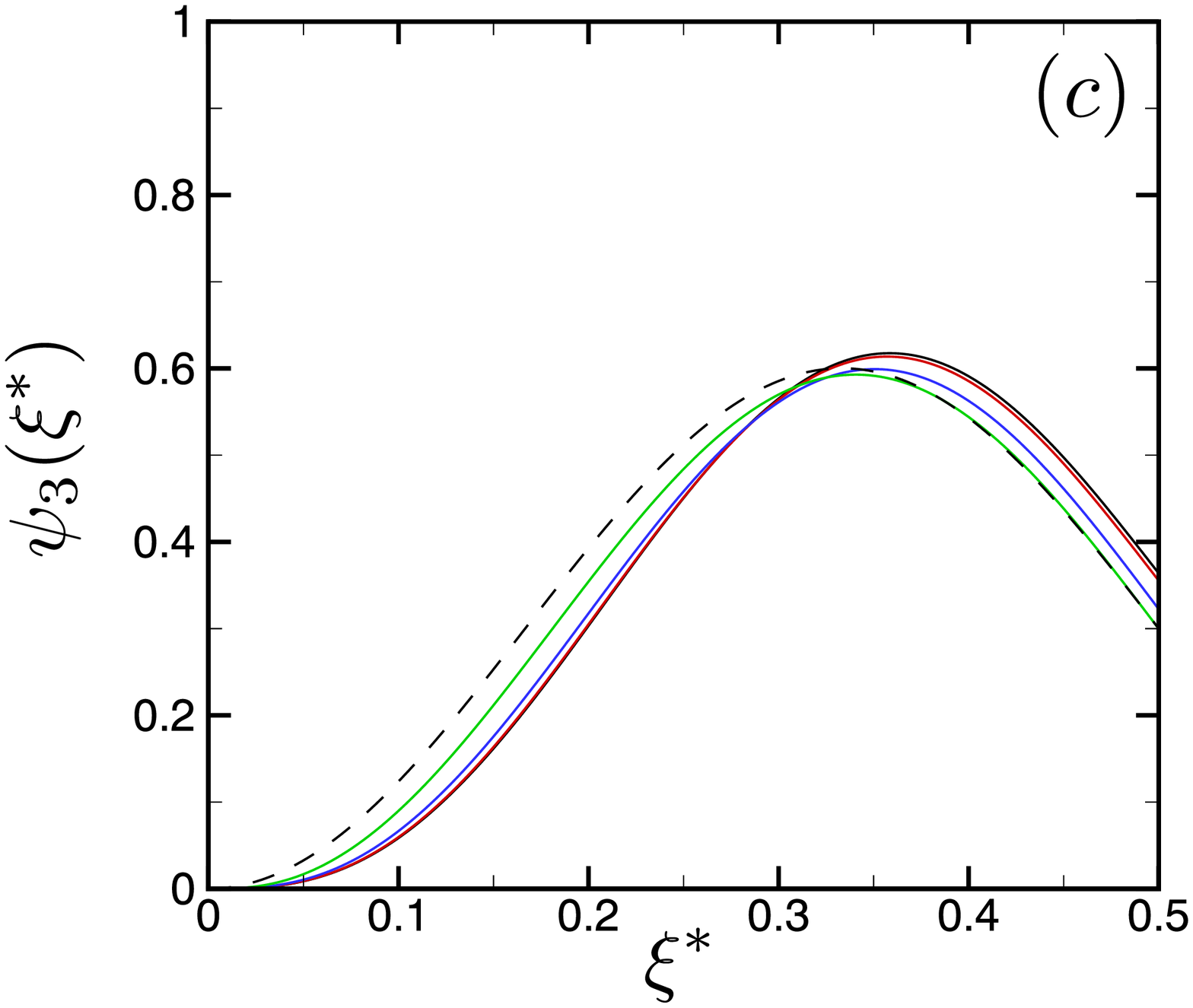} \\ 
\includegraphics[width=2.1in]{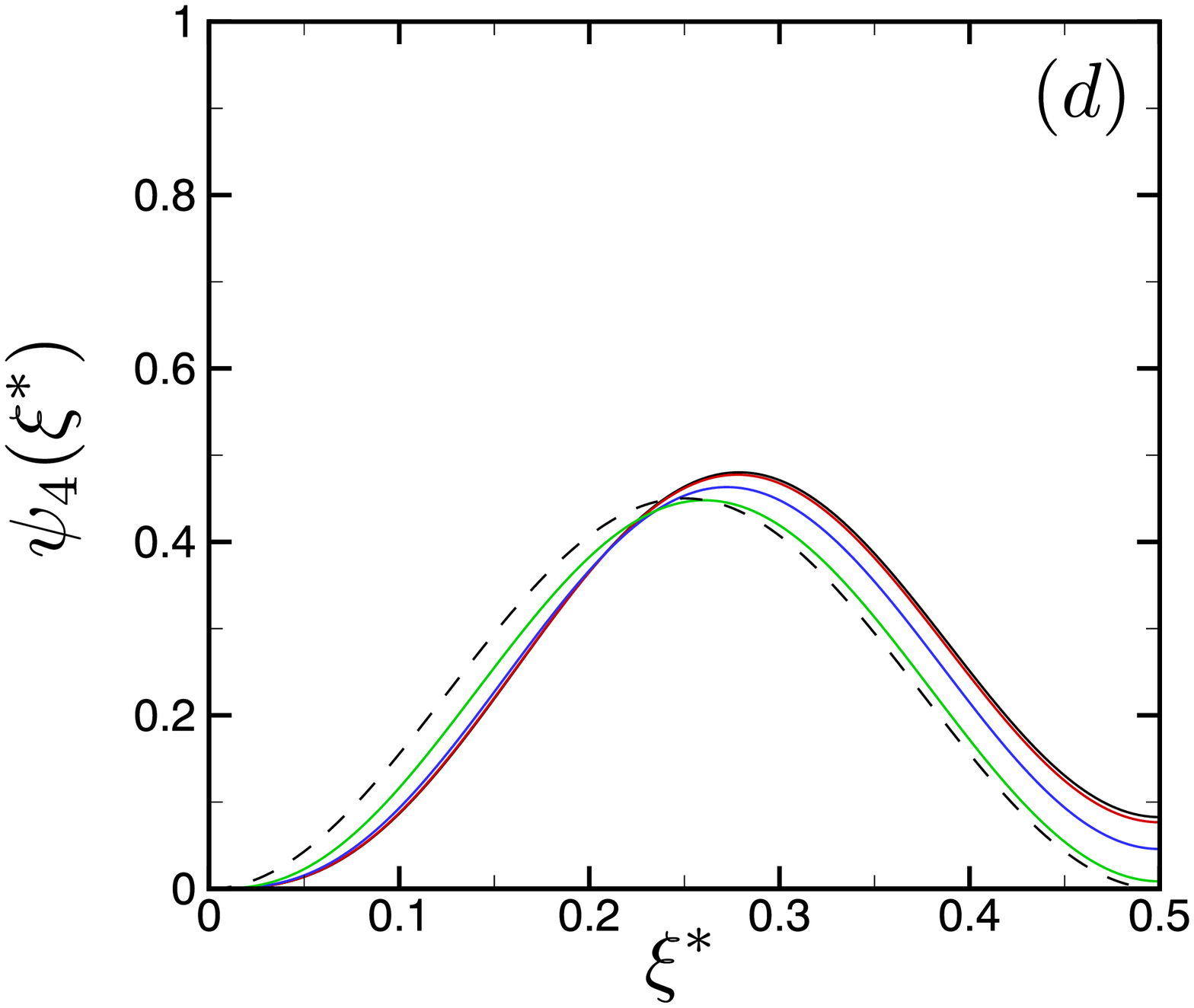}
\includegraphics[width=2.1in]{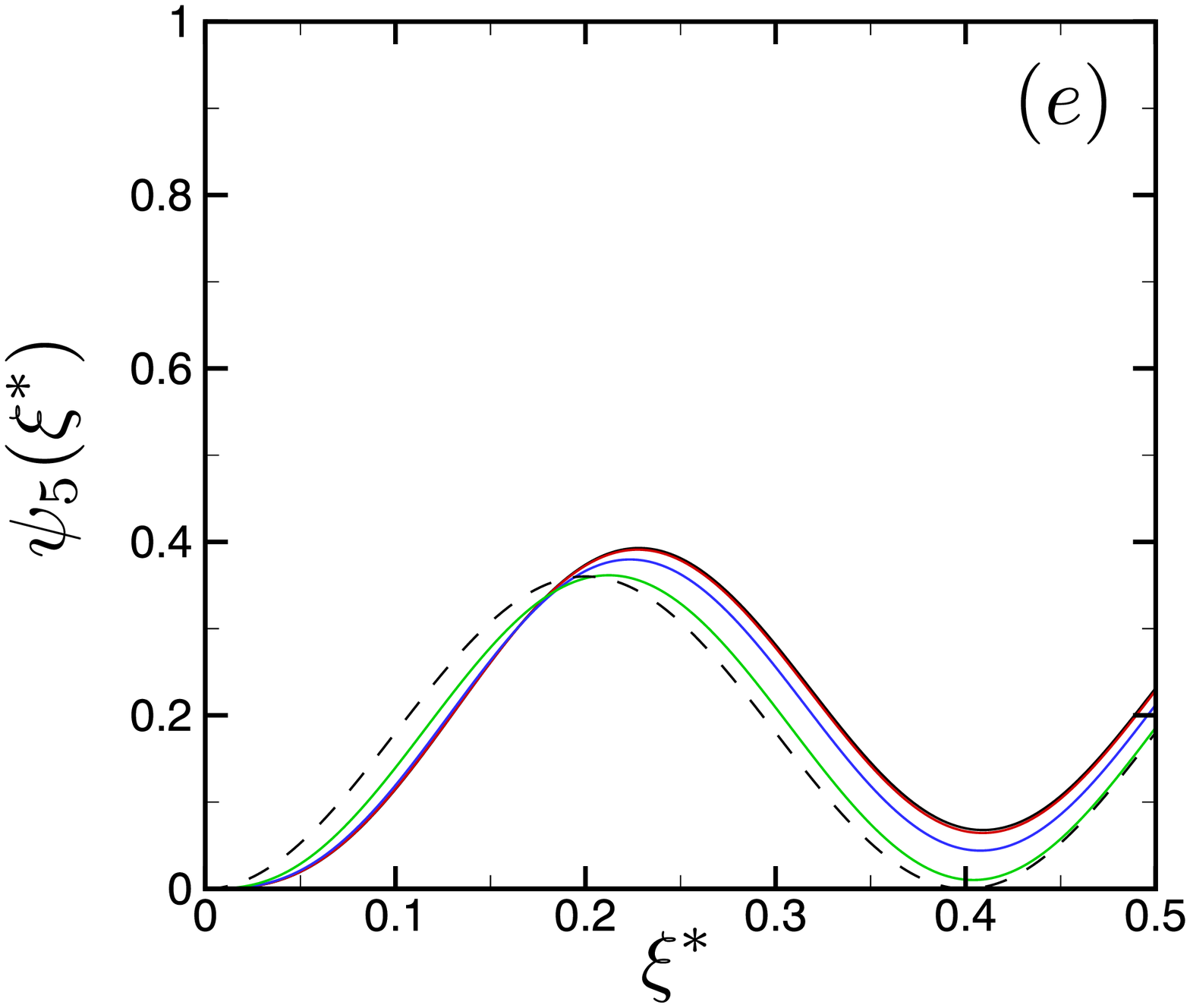}
\end{center}
\caption{The variation of $\psi_n(\xi^*)$ for the first five modes with the distance covered by the applied driving $\xi^*$ for the symmetric drive of the odd modes (a) $n\!=\!1$, (c) $n\!=\!3$ and (e) $n\!=\!5$ or the asymmetric drive of the even modes (b) $n\!=\!2$ and (d) $n\!=\!4$ . The solid curves are for $U\!=\!0$ (black), 10 (red), 100 (blue), 1000 (green) and the dashed curve is for the string. The solid curves approach the string result in order with increasing tension.}
\label{fig:psi-all}
\end{figure}

The variation of $\psi_n$ with $\xi^*$ is shown in Fig.~\ref{fig:psi-all}. The coupling of the spatial driving force with modes 1 through 5 are shown in panels (a)-(e), respectively. The string result is shown as the black dashed line and the solid lines approach the string result in order with increasing tension. In each panel, the 4 solid lines represent results for $U\!=\!0$ (black), 10 (red), 100 (blue), 1000 (green). For the first and second modes, shown in Fig.~\ref{fig:psi-all}(a)-(b), the magnitudes of $\psi_1(\xi^*)$ and $\psi_2(\xi^*)$ increase monotonically with increasing tension. However, for the third mode and higher the coupling with the driving force is non-monotonic with $\xi^*$.

The variation of $\psi_n$ with $\xi^*$ provides immediate insight into the relative heights of the peaks in the amplitude spectrum. Considering only the odd modes, which would be excited by the symmetric drive, this indicates that as $\xi^*$ is increased the coupling with modes 3 and 5 will reach a maximum and will then decay with larger values of $\xi^*$. For example, the relative amplitude of mode 3 will be largest for an applied driving force with $\xi^* \!\approx\! 0.35$ and the relative amplitude of mode 5 will be significantly reduced when $\xi^* \!\approx\! 0.4$. Similar insights can be drawn for the even modes that are driven by an asymmetric driving force.  Knowledge of the $\psi_n(\xi^*)$ could be used in an experiment to tailor the amplitudes of the different modes. Figure~\ref{fig:psi-all} also indicates the influence of the tension on these couplings.
%%%%%%%%%%%%%%%%%%%%%%%%%%%%%%%%%%%%%%%%%%%%%%%%%%%%%%%%%%%%%%%%%%%
%%%%%%%%%%%%%%%%%%%%%%%%%%%%%%%%%%%%%%%%%%%%%%%%%%%%%%%%%%%%%%%%%%%

\subsection{The amplitude of oscillation of a driven beam with tension in a fluid}

\begin{figure}[h]
\begin{center}
\includegraphics[width=3.1in]{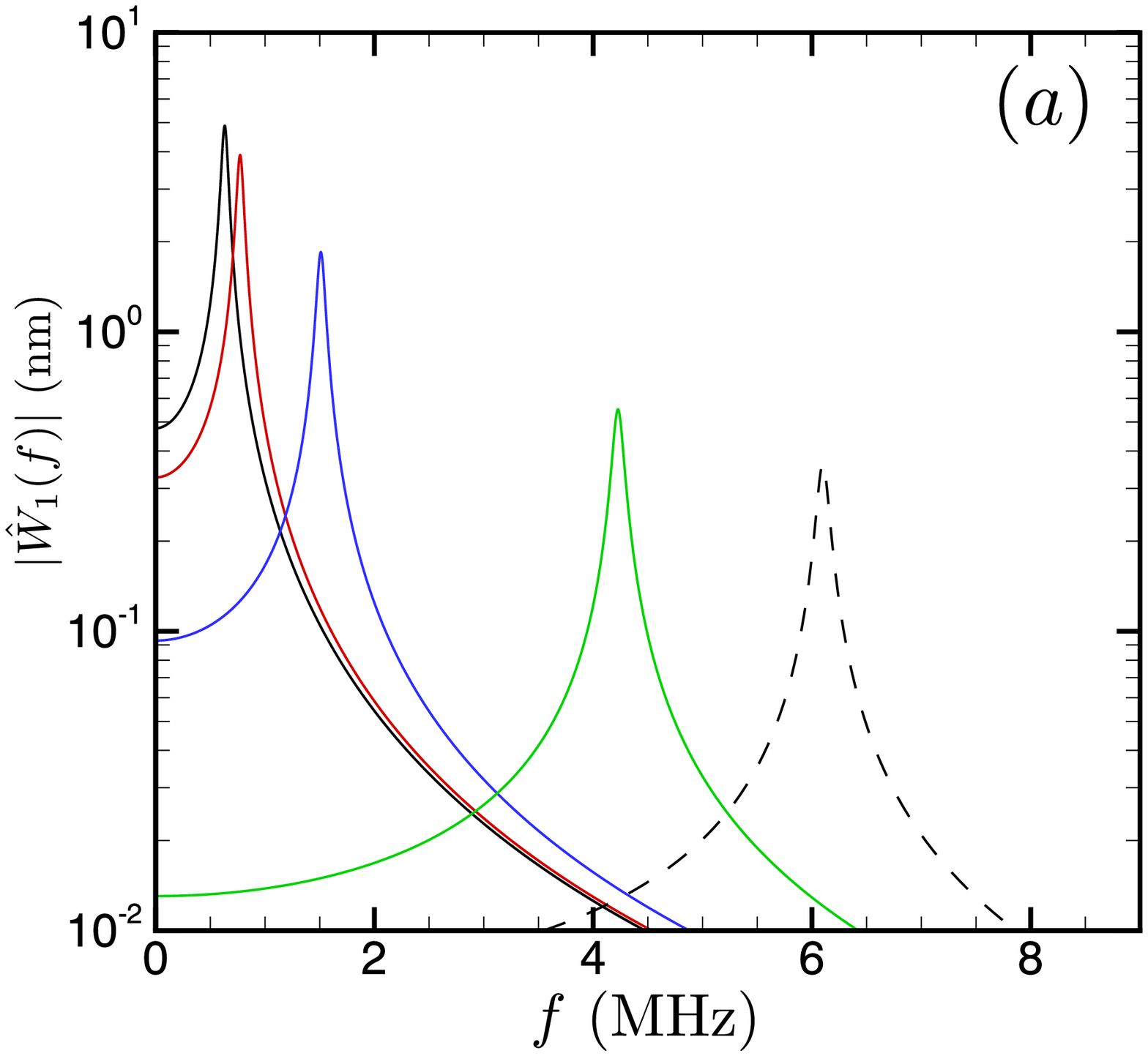}
\includegraphics[width=3.1in]{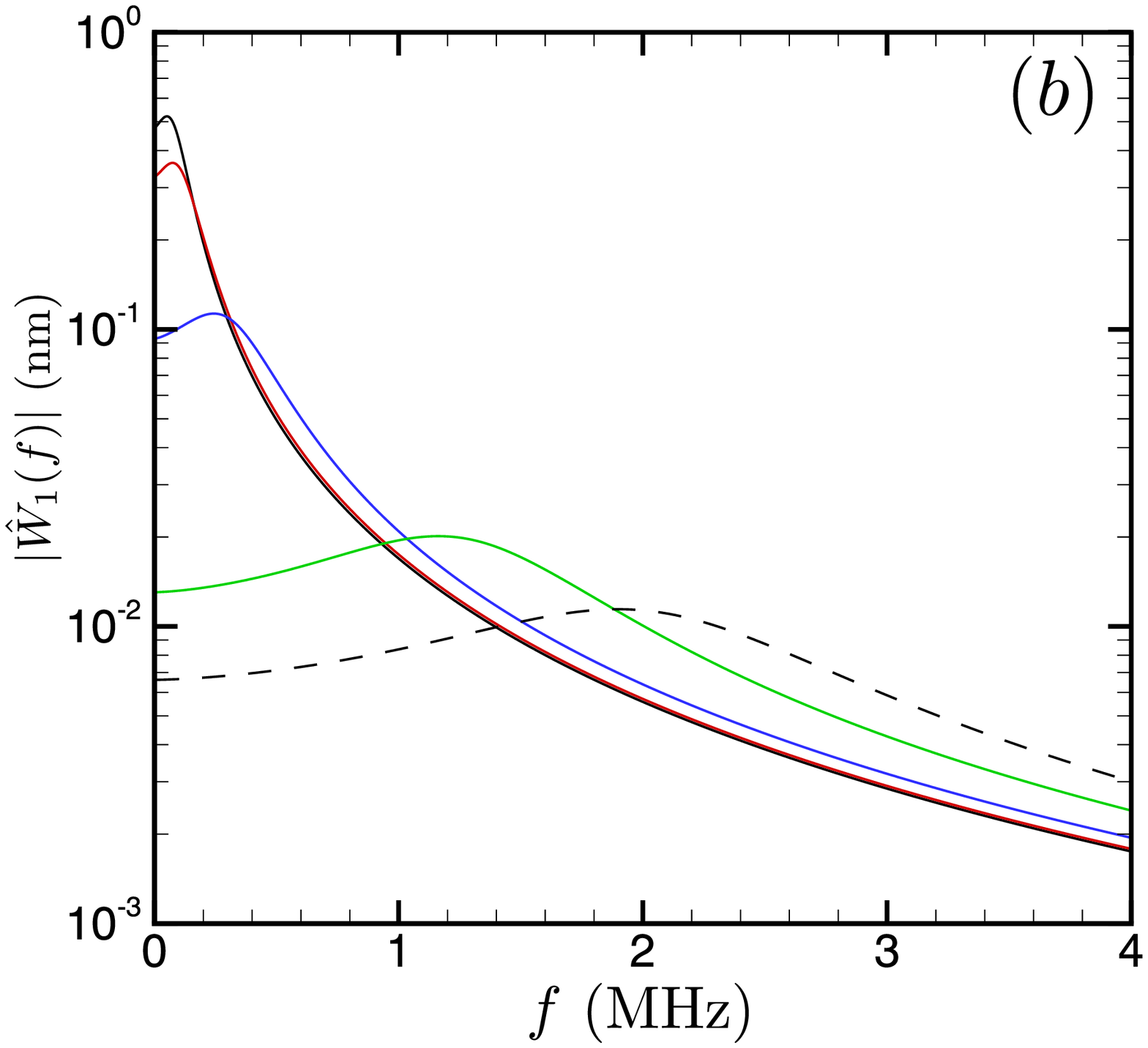}
\end{center}
\caption{The variation of the amplitude of flexural oscillations of the first mode ($n\!=\!1$) with frequency for a beam with increasing tension that is immersed in a fluid. The symmetric drive has a magnitude of $F_0\!=\!2 \!\times\! 10^{-10}$ N and a spatial application of $\xi^*\!=\!0.3$ where the beam is immersed in air~(a) or in water~(b). On each panel the five different curves represent the variation in tension where $U \!=\! 0$ (black), 10 (red), 100 (blue), 1000 (green) and the string (dashed, black). The location of the peak moves towards higher frequency with increasing tension with the string result furthest to the right. These results are obtained by evaluating $|\hat{W}_1(x_0\!=\!1/2,\omega)|$ using Eq.~(\ref{eq:whatnfinal-beam}) for the beam with tension and using Eq.~(\ref{eq:whatnfinal-string}) for the string.} 
\label{fig:air-water-amp}
\end{figure}

The variation of the amplitude of oscillation with frequency for increasing tension is shown in Fig.~\ref{fig:air-water-amp} for the beam given in Table~\ref{table:geom} when it is immersed in air~(a) or water~(b).  Figure~\ref{fig:air-water-amp} shows the amplitude of the fundamental mode, $|\hat{W}(x_0,\omega)|$, measured at $x_0\!=\!1/2$ for an external force of magnitude $F_0\!=\!2\!\times\! 10^{-10}$ N that has been applied over the spatial region given by $\xi^*\!=\!0.3$. The five different curves show the amplitude spectrum of the fundamental mode for varying amounts of tension for $U\!=\!0,10,100,1000$ and a string. As the tension increases, the amplitude of the peak decreases while its frequency increases. The dramatic spreading out of the amplitude spectrum when the beam is immersed in water is evident in Fig.~\ref{fig:air-water-amp}(b). A comparison of the amplitudes in Fig.~\ref{fig:air-water-amp}(a)-(b) also yield an order of magnitude reduction in the amplitude of the beam motion when immersed in water.   
%%%%%%%%%%%%%%%%%%%%%%%%%%%%%%%%%%%%%%%%%%%%%%%%%%%%%%%%%%%%%%%%%%%
%%%%%%%%%%%%%%%%%%%%%%%%%%%%%%%%%%%%%%%%%%%%%%%%%%%%%%%%%%%%%%%%%%%

\begin{figure}[h]
\begin{center}
\includegraphics[width=2.1in]{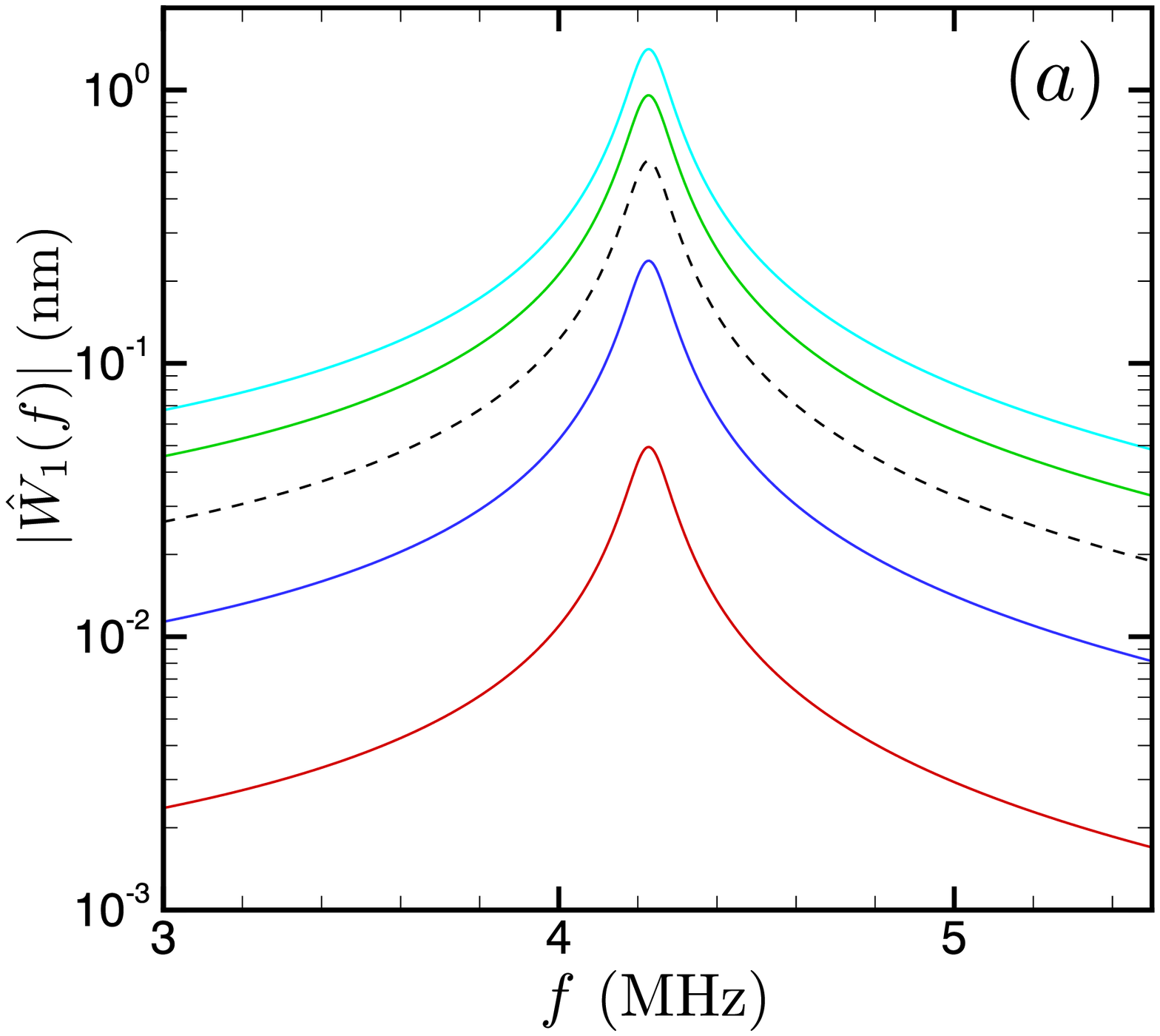}
\includegraphics[width=2.1in]{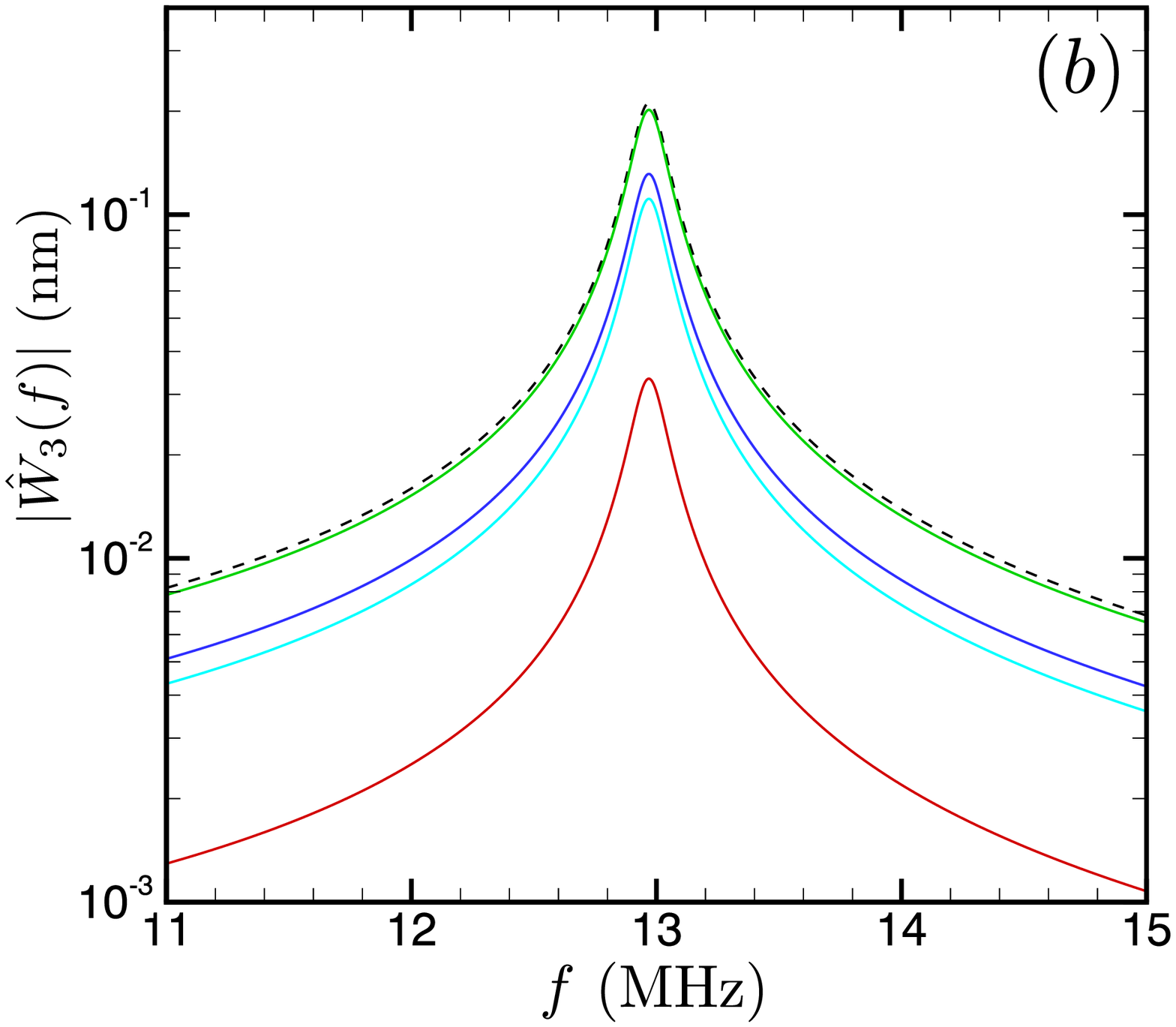}
\includegraphics[width=2.1in]{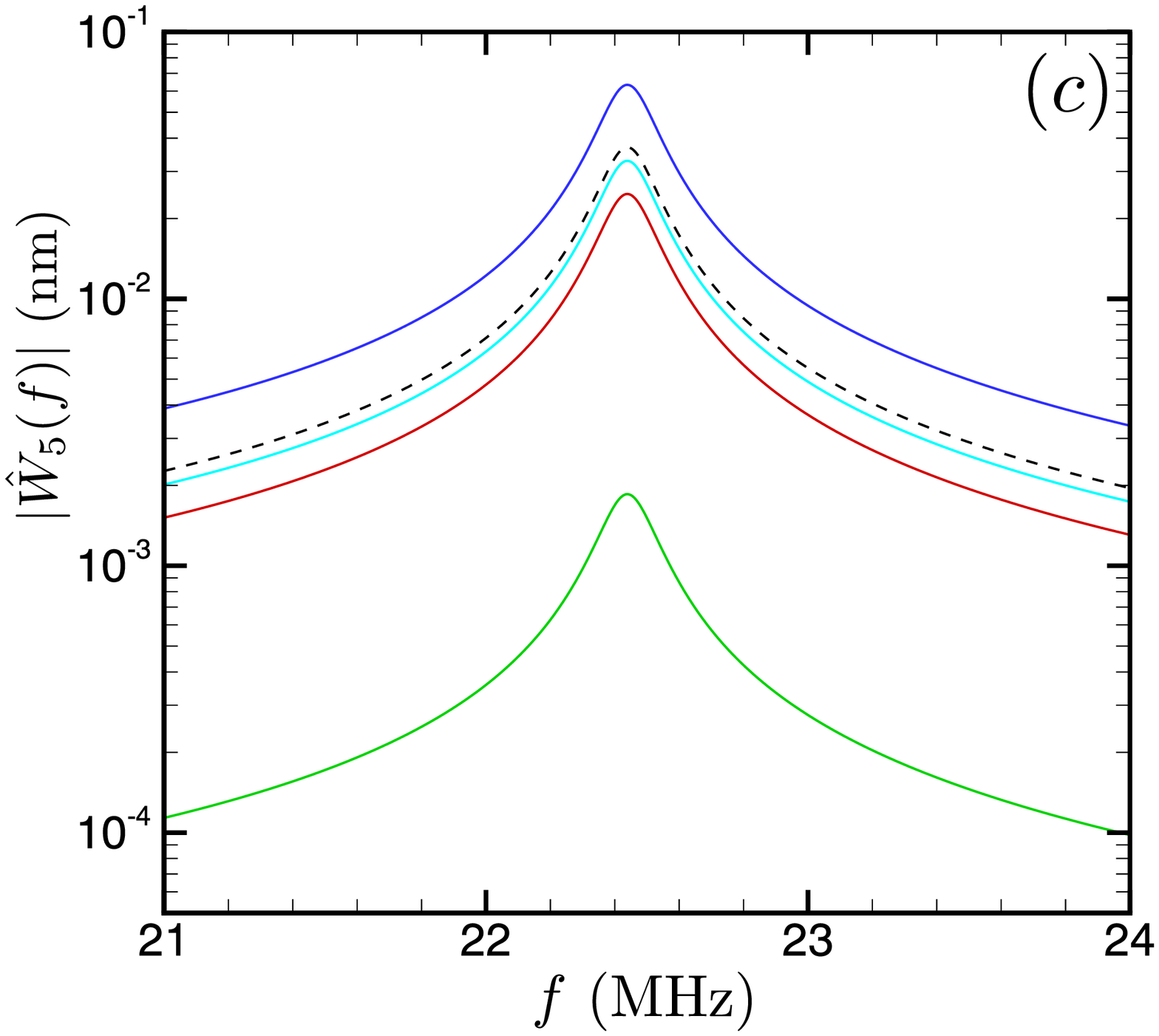}
\end{center}
\caption{The variation of the amplitude of motion of a beam immersed in air under high tension as a function of $\xi^*$ for the odd modes that are driven using a symmetric drive force. The beam is specified in Table~\ref{table:geom}, the tension is $U\!=\!1000$, the magnitude of the drive force is $F_0 = 2 \!\times\! 10^{-10}$ N, and the beam is immersed in air. Each panel includes five curves for $\xi^* \!=\! 0.1$ (red), 0.2 (blue), 0.3 (black, dashed), 0.4 (green), 0.5 (cyan). These results are obtained by evaluating Eq.~(\ref{eq:whatnfinal-beam}) for the odd modes at $x_0\!=\!1/2$. (a) $n\!=\!1$, the peak magnitude increases monotonically with increasing $\xi^*$. (b) $n\!=\!3$, the smallest to largest peak magnitudes occur in the order $\xi^*\!=\!0.1,0.5,0.2,0.4,0.3$. (c) $n\!=\!5$, the smallest to largest peak magnitudes occur in the order $\xi^*\!=\!0.4,0.1,0.5,0.3,0.2$.} 
\label{fig:amp-with-xi-modes135}
\end{figure}

\begin{figure}[h]
\begin{center}
\includegraphics[width=3.1in]{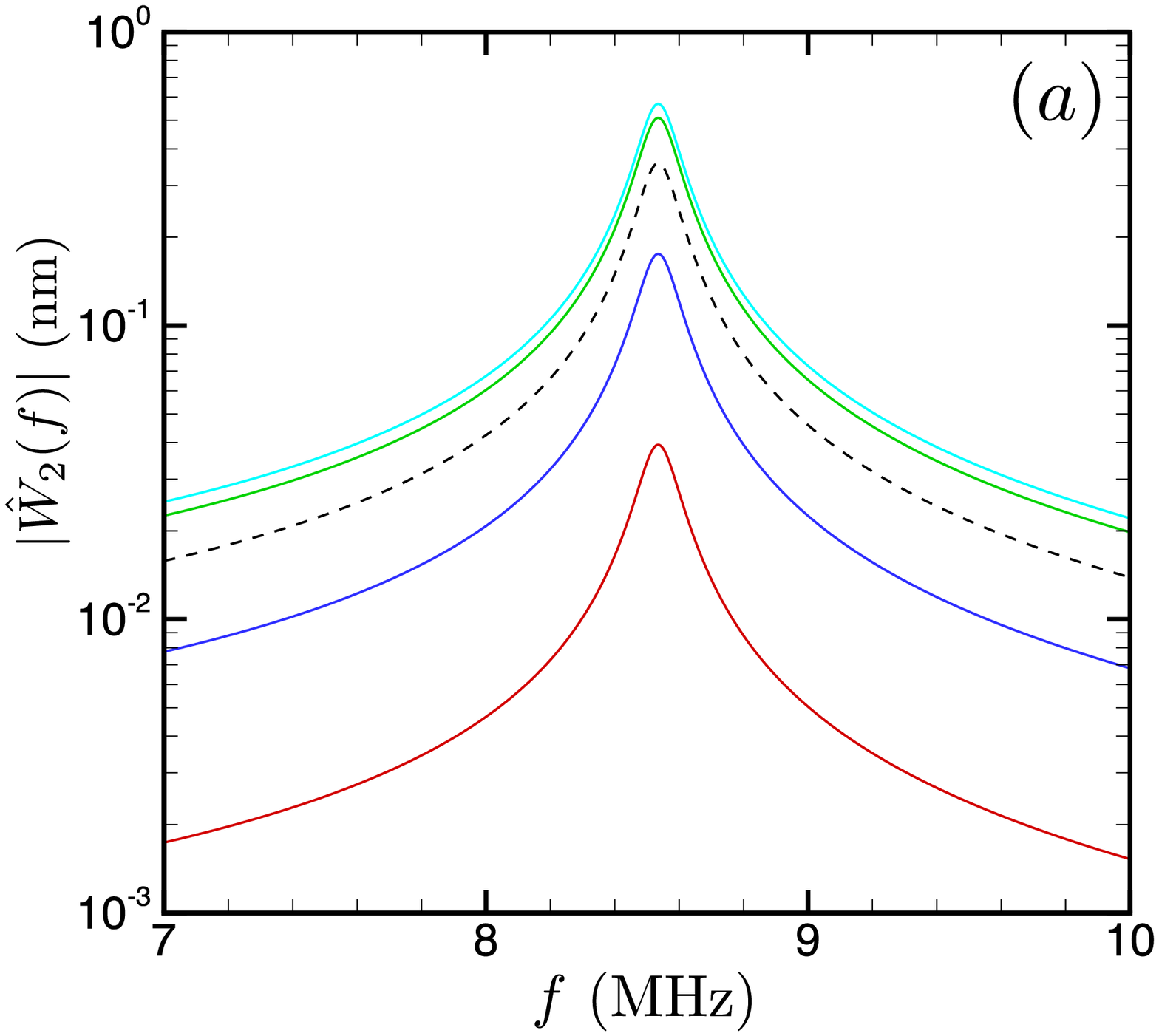}
\includegraphics[width=3.1in]{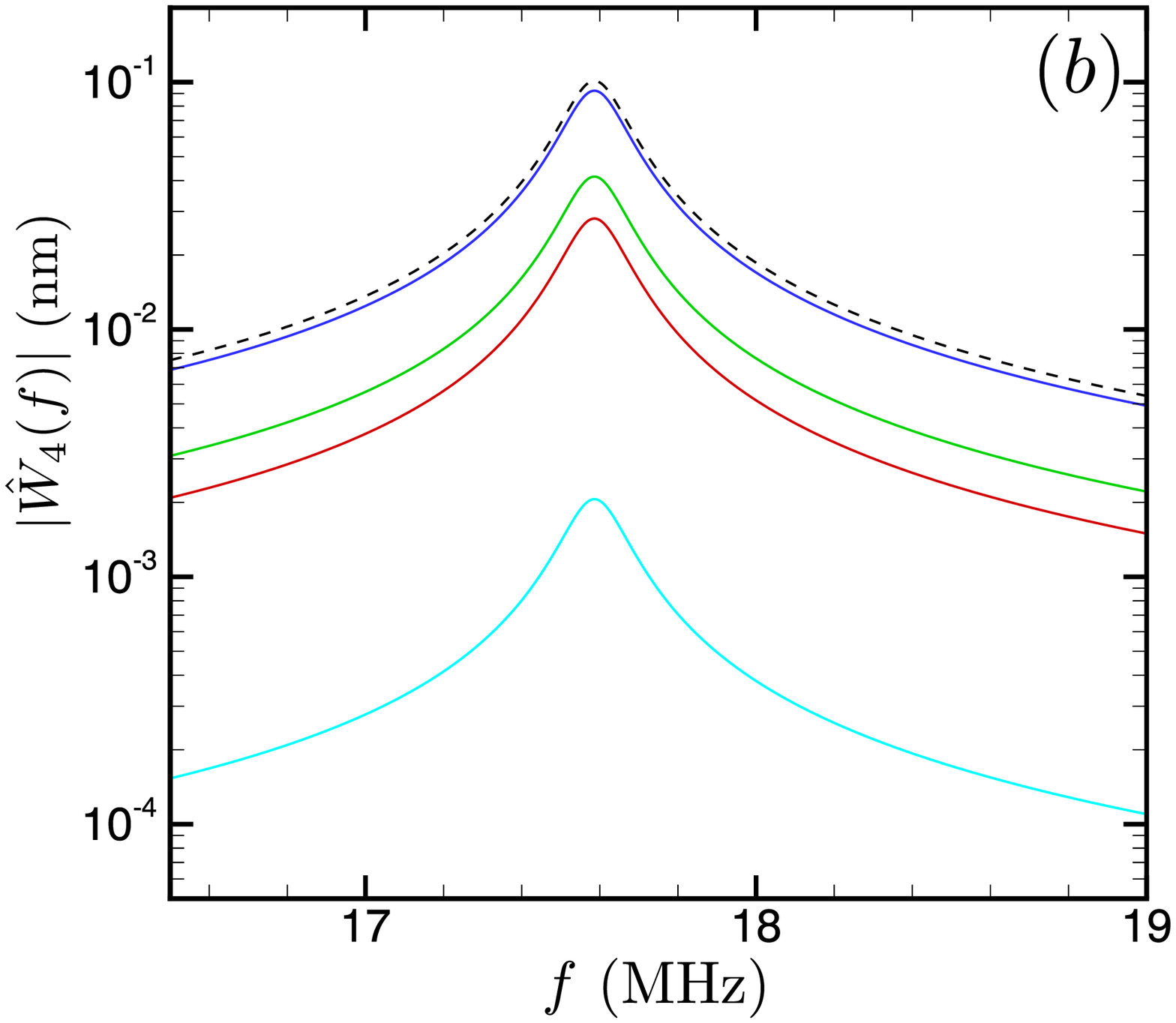}
\end{center}
\caption{The variation of the amplitude of motion of a beam immersed in air under high tension as a function of $\xi^*$ for the even modes that are driven using an asymmetric drive force using the same parameters and conventions as Fig.~\ref{fig:amp-with-xi-modes135}. These results are obtained by evaluating Eq.~(\ref{eq:whatnfinal-beam}) at $x_0\!=\!1/4$ for $n\!=\!2$ and at $x_0\!=\!1/8$ for $n\!=\!4$. (a) Mode $n\!=\!2$, the peak magnitude increases monotonically with increasing $\xi^*$. (b) Mode $n\!=\!4$, the smallest to largest peaks occur in the order $\xi^*\!=\!0.5,0.1,0.4,0.2,0.3$.}
\label{fig:amp-with-xi-modes24}
\end{figure}

The absolute and relative magnitudes of the amplitude peaks depend upon how the external force is applied. In particular, for the situation we explore here, the magnitude of the peaks depend upon the spatial region of the beam that is driven as specified by $\xi^*$.  The variation of the magnitude of the peaks are shown in Fig.~\ref{fig:amp-with-xi-modes135} as a function of $\xi^*$ for a symmetric drive which actuates the odd modes.  The coupling of the drive force with the beam motion is captured by $\psi_n(\xi^*)$ which is directly reflected by the variation of the magnitude of the peaks in Fig.~\ref{fig:amp-with-xi-modes135}. For example, in Fig.~\ref{fig:amp-with-xi-modes135}(a) the magnitudes of the peaks increase monotonically with $\xi^*$ as indicated by the monotonic increase of $\psi_1(\xi^*)$ with $\xi^*$ shown in Fig.~\ref{fig:psi-all}(a). Similarly, Fig.~\ref{fig:amp-with-xi-modes135}(b) shows how the magnitude of the peak increases and then decreases with increasing $\xi^*$ as indicated by the variation of $\psi_3(\xi^*)$ in Fig.~\ref{fig:psi-all}(c). The variation of the magnitude of the peaks of the even modes with $\xi^*$ follow the trends indicated by the appropriate $\psi_n$ and are shown in Fig.~\ref{fig:amp-with-xi-modes24}.
%%%%%%%%%%%%%%%%%%%%%%%%%%%%%%%%%%%%%%%%%%%%%%%%%%%%%%%%%%%%%%%%%%%
%%%%%%%%%%%%%%%%%%%%%%%%%%%%%%%%%%%%%%%%%%%%%%%%%%%%%%%%%%%%%%%%%%%
%%%%%%%%%%%%%%%%%%%%%%%%%%%%%%%%%%%%%%%%%%%%%%%%%%%%%%%%%%%%%%%%%%%
%%%%%%%%%%%%%%%%%%%%%%%%%%%%%%%%%%%%%%%%%%%%%%%%%%%%%%%%%%%%%%%%%%%

\section{Comparison with Experiment}
\label{section:experiment}
\begin{figure}
    \begin{center}
    \includegraphics[width=5in]{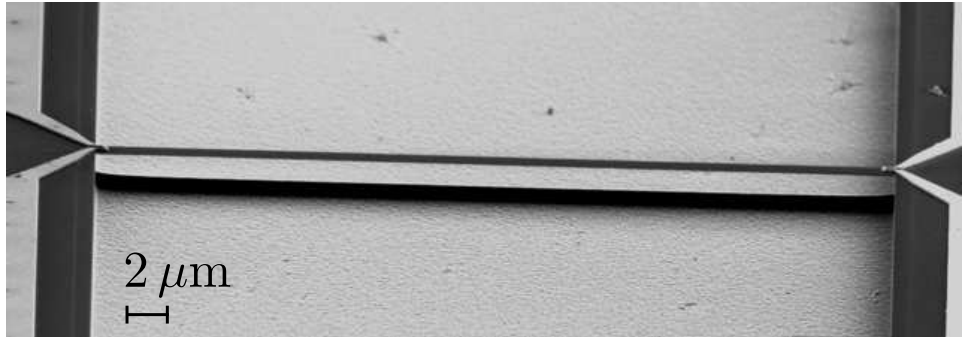}
    \caption{An electron micrograph of the doubly clamped beam used in the experiments. The beam has a length $L\!=\!40 \, \mu$m, width $b \!=\! 0.9 \,  \mu$m,  thickness $h \!=\! 0.1 \, \mu$m, and density $\rho_s \!=\! 2960$ kg/m$^3$. The beam is suspended $2 \, \mu \text{m}$ above the floor. The dark region on the floor below the beam is a residue left over from the fabrication process.}
    \label{fig:experiment}
    \end{center}
\end{figure}

We have performed experiments on a long and thin doubly-clamped beam in the configuration shown in Fig.~\ref{fig:beam}. In the experiments, the beam motion is driven electrothermally near its left and right ends. An electron micrograph of the beam is shown in Fig.~\ref{fig:experiment}. The experimental approach has been described in detail elsewhere~\cite{ari:2021,ti:2021} and we provide only the essential details here.

The geometry and density for this beam are specified in Table~\ref{table:geom}.  The value listed for the beam density in Table~\ref{table:geom} is the experimentally measured value. As a result, we will consider the density of the beam a known value. The Young's modulus $E$ in Table~\ref{table:geom}, on the other hand, is simply a nominal value for silicon nitride~\cite{ari:2021}. As shown in Fig.~\ref{fig:experiment}, the entire beam is suspended in a cavity. The distance from the beam to the substrate below is approximately 2 $\mu$m. The beam is under high tension, $U \!\gg\! 1$, as a result of the fabrication process. We quantify a value for $U$ in the discussion below. We have conducted experiments on this beam when it is placed in vacuum, air, or water.
%%%%%%%%%%%%%%%%%%%%%%%%%%%%%%%%%%%%%%%%%%%%%%%%%%%%%%%%%%%%%%%%%%%
%%%%%%%%%%%%%%%%%%%%%%%%%%%%%%%%%%%%%%%%%%%%%%%%%%%%%%%%%%%%%%%%%%%

The upper surface of the beam contains two U-shaped gold electrodes near the sidewalls, with one located at the left side and the other located at the right side, that are used for the electrothermal drive. The gold electrodes are evident in Fig.~\ref{fig:experiment} by their lighter color near the locations where the beam attaches to the sidewalls. When a sinusoidally varying current passes through the gold electrodes the beam is locally heated due to Joule heating. The heating causes the beam to expand which results in beam deflections since the beam is fixed at either end. The odd modes of the beam are actuated when the current in the two electrodes is in-phase (symmetric drive) and the even modes are actuated when the current is 90$^\circ$ out-of-phase (asymmetric drive). The deflection of the beam is measured optically at location $x_0$ which is typically chosen to correspond with an anti-node of the beam mode that is being measured.

We next describe the procedure used for comparing the theoretical predictions with the experimental measurements. The constants $U$ and $F_T$ are required in the theoretical expressions to describe the amount of tension in the beam.  In addition, we will introduce an effective length $L^*$ of the beam in order to quantitatively describe the magnitudes of the natural frequencies for the  conditions of the experiment. The constants $\xi^*$ and $F_0$ are required to describe the spatially varying drive force that is applied. Detailed measurements of the tension and the drive force are difficult to obtain directly and, in our analysis, we determine their experimental values indirectly using only measured frequencies, amplitudes, and the density of the beam which are typically more accessible.

The values of $U$, $L^*$, and $F_T$ are determined using measurements of the beam's natural frequencies $\omega_{n,exp}$ in a vacuum where $n \!=\! 1, 2, \ldots N$. We have used $N\!=\!5$  but the   approach remains valid if more or fewer natural frequencies are used.

\begin{enumerate}
    \item Determine the tension parameter $U$. The value of $U$ determines the nondimensional natural frequencies $\Omega_n$ and their spacing in the frequency domain as the roots of Eq.~(\ref{eq:characteristic-equation}). We determine $U$ as the value which minimizes the total error $E_U$ in the ratio of each mode with the fundamental frequency when compared with the experimentally measured values. Specifically, $U$ is chosen as the value which minimizes $E_U$ where
    \begin{equation}
    E_U = \sum_{n=2}^N \left(\frac{\Omega_n}{\Omega_1} - \frac{\omega_{n,exp}}{\omega_{1,exp}}\right)^2
    \label{eq:ufit}
    \end{equation}
    and the subscript exp indicates the experimentally measured natural frequencies. This step determines $U$ and $\Omega_n$.  For large values of the tension it is important to measure $\omega_n/\omega_1$ in vacuum, and to not approximate their values using measurements in air which can lead to a significant error in the determination of $U$.
    
    \item Determine the effective length $L^*$. The effective length is chosen to obtain quantitative agreement with the dimensional natural frequencies measured in experiment. For each mode $n$, an effective length can be computed using Eq.~(\ref{eq:Omega}) and requiring $\omega_n \!=\! \omega_{n,exp}$. We define the effective length $L^*$ as the mean of these values which can be expressed as  
    \begin{equation}
        L^* = \frac{1}{N} \sum_{n=1}^N \left(\frac{E I\Omega_n^2}{\mu\omega_{n,exp}^2}\right)^{1/4}
        \label{eq:lfit}
    \end{equation}
where $\Omega_n$ is the nondimensional natural frequency of the beam with tension.
    \item Determine the tension force $F_T$.  The tension force is found by evaluating Eq.~(\ref{eq:U}) to yield
    \begin{equation}
        F_T = \frac{2 E I U}{L^{*2}}.
        \label{eq:ft-using-estar}
        \end{equation}
\end{enumerate}

Applying this procedure to the beam used in the experiments yields the values given in the first row of Table~\ref{table:properties}.  The large value of the tension parameter, $U\!=\!4538$, indicates that the beam is under very high tension which immediately suggests that a string description may be useful. The effective length $L^*$ is larger than the length of the beam $L$. We emphasize that the effective length is also being used to account for features of the experiment that has not been explicitly included in the theoretical description. This includes the presence of the ledge which the beam is attached to at its edges, the gold electrodes that are used for the electrothermal drive, and any material inhomogeneities that may be present. Lastly, we note that we have used a nominal value for the Young's modulus, $E \!=\! 300$ GPa.  Using this approach results in theoretical predictions of the first five natural frequencies of the beam with tension that have errors of less than 0.3\% when compared with the experimentally measured values. The last row of Table~\ref{table:properties} includes the value of $F_T$ that is for the string description. In this case, $\omega_{1,exp}$ is used to determine $F_T$ as $F_T \!=\! c^2 \mu$ which can be expressed as
\begin{equation}
    F_T = \omega_{1,exp}^2 \frac{L^2}{\pi^2} \mu.
    \label{eq:FT-string}
\end{equation}
%%%%%%%%%%%%%%%%%%%%%%%%%%%%%
\begin{table}[h!]
\begin{center}
\begin{tabular}{ c c c c} 
\vspace{-0.2cm}
theory  & $U$   & $L^*$     & $F_T$     \\
        &       & ($\mu$m)  & ($\mu$N)  \\ \hline \hline
beam    & 4538  & 48.0      & 88.6 \\
string  & --    & --        & 65.0\\
\end{tabular}
\end{center}
\caption{Parameters required in the theoretical description that are determined using experimental measurements of the natural frequencies. Equation~(\ref{eq:ufit}) is used for the tension parameter $U$,  Eq.~(\ref{eq:lfit}) is used for the effective length $L^*$, and Eq.~(\ref{eq:ft-using-estar}) is used for the tension force $F_T$. For the string, $F_T$ is determined by Eq.~(\ref{eq:FT-string}). The geometry and material properties of the beam are given in Table~\ref{table:geom}.}
\label{table:properties}
\end{table}
%%%%%%%%%%%%%%%%%%%%%%%%%%%%%

The theoretical prediction also requires values of $\xi^*$ and $F_0$ which are used to describe the spatially varying drive force that is applied. In the experiments, the magnitude of the driving was set differently depending upon if the beam was in vacuum, air,  or water. The driving magnitude in the experiments was then maintained at a constant value for the symmetric and asymmetric drive measurements.  In order to account for the differences in the symmetric and asymmetric driving of the experiments we determined values of $\xi^*$ and $F_0$ for each measurement that we used in our comparison between theory and experiment.  We point out that the  symmetric and asymmetric driving measurements are treated separately to yield values of $\xi^*$ and $F_0$ that we use in our theoretical predictions. For each measurement, we have used the following steps to determine $\xi^*$ and $F_0$.
\begin{enumerate}
    \item Determine the spatial application of the driving force $\xi^*$.  The value of $\xi^*$ is used to determine the ratio of the amplitudes measured at their peak frequency. For the odd modes, driven by the symmetric drive, $\xi^*$ is determined as the value which minimizes the total error, $E_{amp}$, in the experimentally measured ratios of the amplitude of the higher modes at their peaks to the magnitude of the first mode at its peak. All odd modes are measured at $x_0=1/2$. Therefore, for the odd modes, $\xi^*$ is the value that minimizes $E_{amp}$ where 
    \begin{equation}
        E_{amp} = \left(\frac{|\hat{W}_3(x_0,\omega_{3,f})|}{|\hat{W}_1(x_0,\omega_{1,f})|} - \frac{|\hat{W}_3(x_0,\omega_{3,f})|_{exp}}{|\hat{W}_1(x_0,\omega_{1,f})|_{exp}}\right)^2 + \left(\frac{|\hat{W}_5(x_0,\omega_{5,f})|}{|\hat{W}_1(x_0,\omega_{1,f})|} - \frac{|\hat{W}_5(x_0,\omega_{5,f})|_{exp}}{|\hat{W}_1(x_0,\omega_{1,f})|_{exp}}\right)^2
    \end{equation}
    Similarly, for the even mode experiments $\xi^*$ is the value that minimizes 
        \begin{equation}
        E_{amp} = \left(\frac{|\hat{W}_4(x_0,\omega_{4,f})|}{|\hat{W}_2(x_0,\omega_{2,f})|} - \frac{|\hat{W}_4(x_0,\omega_{4,f})|_{exp}}{|\hat{W}_2(x_0,\omega_{2,f})|_{exp}}\right)^2
        \end{equation}
        where $x_0\!=\!1/4$ for mode 2 and $x_0\!=\!1/8$ for mode 4. 
    \item Determine the magnitude of the external force $F_0$. The value of $F_0$ is used to set the overall magnitude of the amplitudes of the modes at their peak frequency. Note that is the previous step it was the ratio of the magnitudes that was determined and not the overall magnitude.  $F_0$ is determined as the value required for the amplitude of the first mode to agree with the experimentally measured value. For the odd mode experiments, this can be expressed as
    \begin{equation}
        |\hat{W}_1(x_0,\omega_{1,f})| = |\hat{W}_1(x_0,\omega_{1,f})|_{exp}
    \end{equation}
    where $x_0\!=\!1/2$. For the even mode experiments this becomes
    \begin{equation}
        |\hat{W}_2(x_0,\omega_{2,f})| = |\hat{W}_2(x_0,\omega_{2,f})|_{exp}
    \end{equation}
    where $x_0\!=\!1/4$. Combined with the choice of $\xi^*$ in the previous step, this ensures that peak magnitude of the amplitude of the first mode agrees with experimental measurements.
\end{enumerate}
%%%%%%%%%%%%%%%%%%%%%%%%%%%%%
\begin{table}[h!]
\begin{center}
\begin{tabular}{ c c c c c c } 
\vspace{-0.2cm}
theory & modes  & $\xi^*$    & $F_0$ & $\xi^*$   & $F_0$ \\ \vspace{-0.2cm}
      &         & air        & air   & water     & water  \\
      &         &            & (nN)  &           & (nN) \\ \hline \hline 
beam  & odd     & 0.275      & 0.13  & 0.26      & 1.69 \\ 
beam  & even    & 0.225      & 0.15  & 0.195     & 1.69 \\ 
string & odd     & 0.275      & 0.15  & 0.26      & 1.96 \\ 
string & even    & 0.225      & 0.18  & 0.195     & 1.96 \\ 
\end{tabular}
\end{center}
\caption{Parameters used in the theoretical description of the external driving that are determined using experimental measurements. $\xi^*$ specifies the spatial region of the beam where the driving force is applied and $F_0$ is the magnitude of the driving force. $\xi^*$ and $F_0$ depend on the experiment as indicated by the fluid used. These parameters are determined in order from left to right as described in Section~\ref{section:experiment}. Rows 1-2 are for the beam theory with tension and rows 3-4 are for the string theory.}
\label{table:forcing-parameters}
\end{table}
%%%%%%%%%%%%%%%%%%%%%%%%%%%%%

The values of $\xi^*$ and $F_0$ are given in Table~\ref{table:forcing-parameters}. The first two rows shows the parameters for the beam theory and the last two rows show the parameters for the string theory. The electrothermal driving is tailored individually for each experiment therefore our values of $\xi^*$ and $F_0$ also vary with each experiment.  The spatial extent of the applied force $\xi^*$ for the odd mode experiments, for both air and water, encompasses approximately 40-55\% of the beam's upper surface when the driving at both ends of the beam are included. This is a much larger distance than the physical region of the beam in contact with the gold electrodes as shown in Fig.~\ref{fig:experiment}. This illustrates the complexity of the electrothermal drive. Our model suggests that to represent the electrothermal drive as a uniformly applied harmonic force requires a much larger spatial application than what is covered by the electrodes in order to yield the measured amplitudes of deflection. Our value of $\xi^*$ provides a measure of the effective spatial extent of this driving when represented as a harmonic force with a constant magnitude. It would be an interesting study to explore the details of the electrothermal drive in depth. Our intention here is not to focus on the physics of one particular driving mechanism, but rather to explore in general how a spatially varying drive affects the dynamics of a beam immersed in a fluid over a wide range of tension. 

The magnitude of $F_0$ is more than an order of magnitude larger for the experiment in water when compared to the experiment in air. In general, the voltages used for the electrothermal driving are set to larger values when a more viscous and denser fluid is used in order to achieve a desired magnitude of fluctuations. 

\subsection{The beam immersed in air}

We first present a comparison of the theoretical prediction with experimental measurement for the case when the beam is immersed in atmospheric air. It is important to emphasize that the hydrodynamic function $\Gamma(\omega)$, introduced in Eq.~(\ref{eq:fluid-force}), assumes that the fluid obeys the continuum Newtonian approximation for the length and time scales under consideration~\cite{rosenhead:1963,sader:1998}. However, for nanoscale devices in air this assumption can be violated~\cite{PhysRevLett.118.074505}. The important nondimensional numbers that provide insight into this issue are the Knudsen number $\text{Kn}$ and the Weissenberg number $\text{Wi}$. The Knudsen number can be estimated as the ratio of length scales  $\text{Kn} \!=\! \lambda /b$ where $\lambda$ is the mean-free path of air and $b$ is the beam width. The Weissenberg number is the ratio of time scales  $\text{Wi} \!=\! \tau/\omega^{-1}$ where $\tau$ is the relaxation time and $\omega^{-1}$ is the inverse of the oscillation frequency of interest for the beam. Using $\lambda \approx 100$ nm for atmospheric air yields $\text{Kn} \!\approx\! 0.1$ and using a relaxation time\cite{PhysRevLett.118.074505} of $\tau \!\approx\! 8 \times 10^{-10}$ s yields $\text{Wi} \!\approx\! 0.1$ if we use the frequency of the fifth mode of the beam.  The scaling shown in Ref.~\cite{PhysRevLett.118.074505} suggests that the continuum approximation for the hydrodynamic function is satisfactory when $\text{Kn} \!+\! \text{Wi} \!\lesssim\! 1$. In light of this, we expect that the continuum description will remain valid for the range of measurements we have conducted in air.

\begin{figure}[h]
\begin{center}
\includegraphics[width=2.1in]{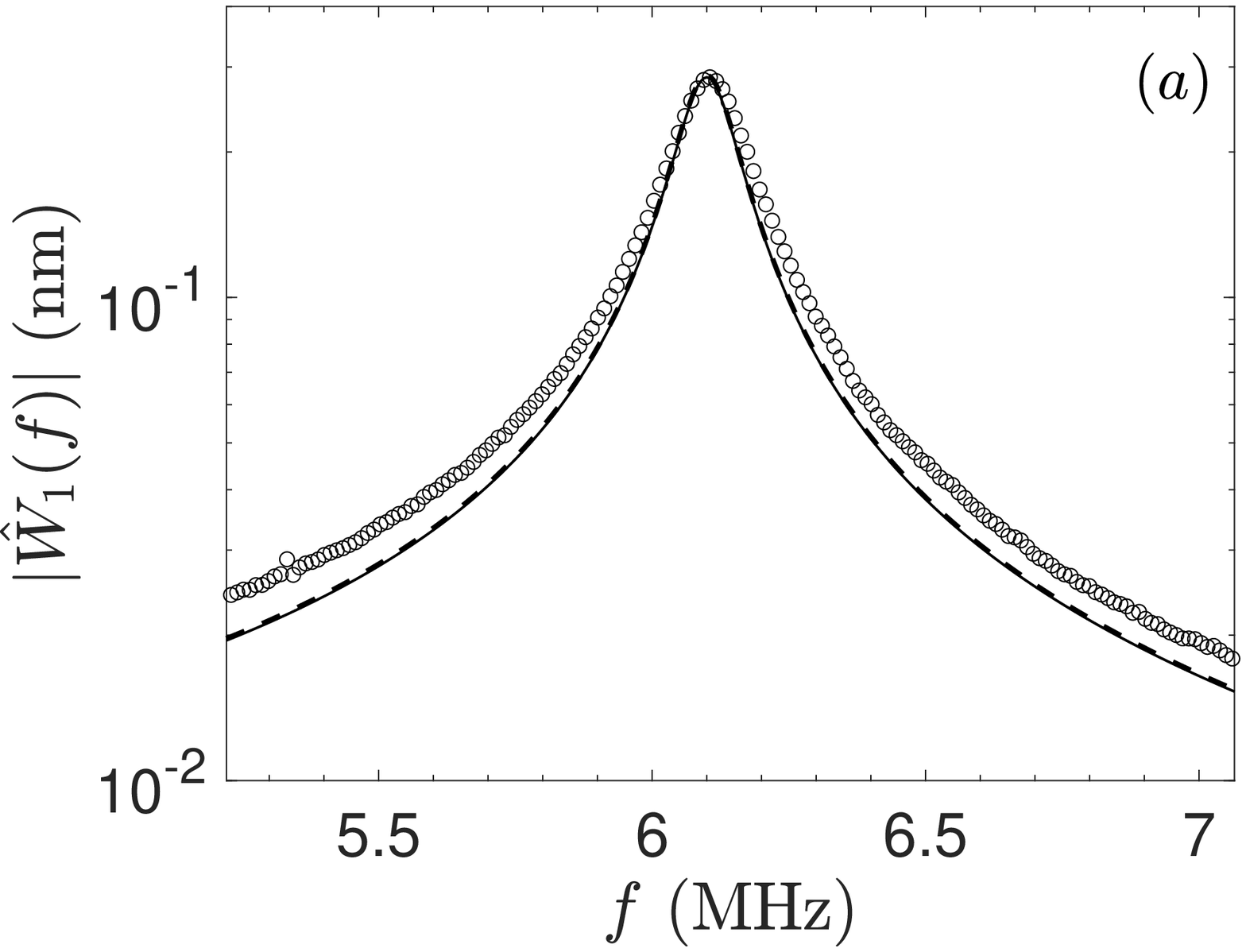}
\includegraphics[width=2.1in]{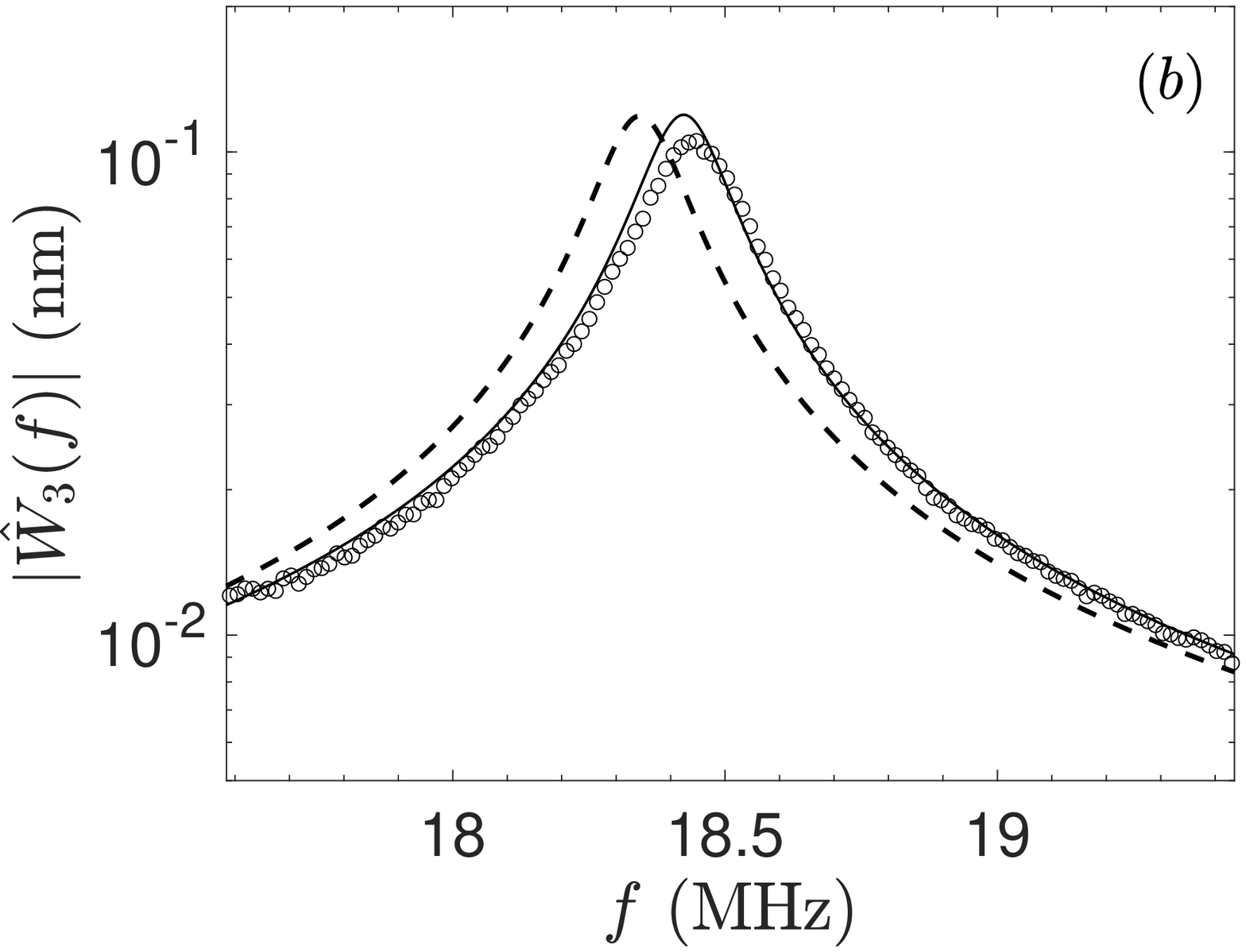} 
\includegraphics[width=2.1in]{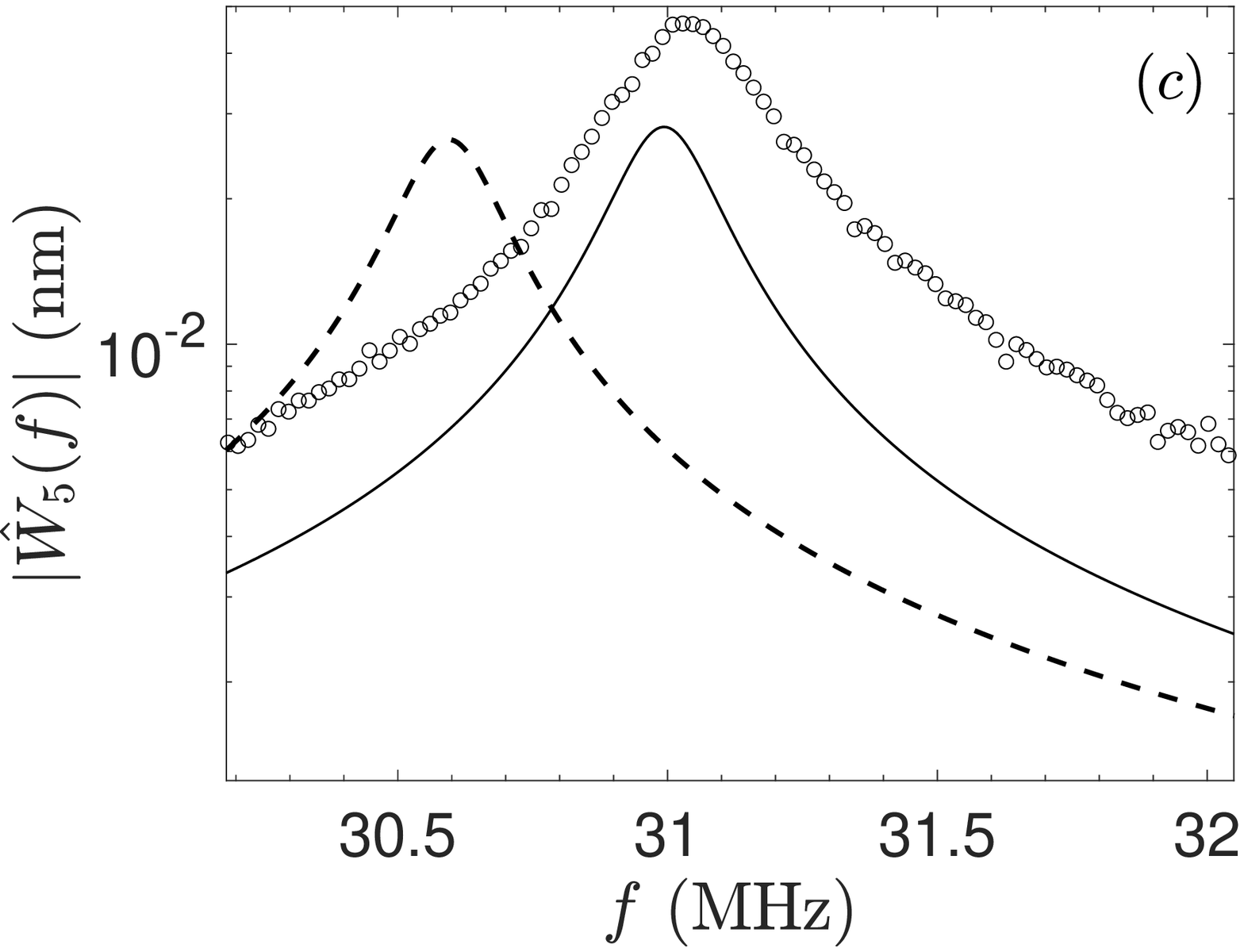}
\end{center}
\caption{The frequency dependent amplitude of oscillation for the odd modes when the beam is immersed in air. Open circles are experimental measurements, the solid lines are the theoretical prediction using beam theory given by Eq.~(\ref{eq:whatnfinal-beam}) with $n\!=\!1,3,5$, and the dashed lines are the string prediction given by Eq.~(\ref{eq:whatnfinal-string}) with  $n\!=\!1,3,5$. The beam motion is evaluated at $x_0\!=\!1/2$ and a symmetric driving force is used. See Table~\ref{table:properties} for detailed information about the specific properties. (a)-(c) Show results for modes $n\!=\!1$, 3, and 5 respectively.}
\label{fig:theory-experiment-air-odd}
\end{figure}

The amplitude spectra are shown in Fig.~\ref{fig:theory-experiment-air-odd} for the odd modes. The parameters used in the beam theory are given in Table~\ref{table:properties}. Figure~\ref{fig:theory-experiment-air-odd}(a) shows the comparison for the fundamental mode where the solid line is the theoretical prediction of the beam theory given by Eq.~(\ref{eq:whatnfinal-beam}) using $n\!=\!1$ and the open circles are the experimental measurements. The experiment uses the symmetric electrothermal drive and the motion of the beam is measured at $x_0\!=\!1/2$ for all of the odd modes. We note that for air it is sufficient to represent the amplitude spectrum using only the mode of interest, as opposed to the full mode expansion, since the peaks of the different modes are sharp and well separated in frequency.  Also included is the prediction using the string description, given by Eq.~(\ref{eq:whatnfinal-string}) using $n \!=\!1$, which is represented as the dashed line. The dashed line is nearly indistinguishable from the solid line indicating excellent agreement.
%%%%%%%%%%%%%%%%%%%%%%%%%%%%%%%%%%%%%
%%%%%%%%%%%%%%%%%%%%%%%%%%%%%%%%%%%%%

The comparison between theory and experiment for mode 3 is shown in Fig.~\ref{fig:theory-experiment-air-odd}(b) where the agreement between the beam theory and the experiment is excellent. The string approximation now shows some error in the location of the peak frequency as expected since the string approximation is the limit of infinite $U$ and the actual beam has a large but finite value of $U$.  We would like to highlight that the peak of mode 3 for the string description is less than the frequency of the peak for the beam theory.  This is because we have set $F_T$ for the string description to yield the experimentally measured fundamental frequency as shown in Fig.~\ref{fig:theory-experiment-air-odd}(a).  The spacing of the peaks in frequency space is smaller for the string than for the beam with tension as shown in Fig.~\ref{fig:freq}(b). Therefore the peak frequency of the string is lower in Fig.~\ref{fig:theory-experiment-air-odd}(b) than that of the beam. In fact, for this reason, all of the string predictions that we will show will include a shift toward lower frequencies for this reason.  However, we emphasize that the string approximation represents mode 3 quite accurately despite this shift in frequency. 

Figure~\ref{fig:theory-experiment-air-odd}(c) illustrates the comparison for mode 5 where the agreement with the beam theory is good for the shape and location of the amplitude spectrum however it significantly under predicts the magnitude.  The error in the magnitude of the motion of mode 5 is expected to be due to the coupling of the electrothermal drive to mode 5 which is stronger than what is described by our model. The string approximation is included as the dashed line which again is similar to the result from the beam theory with a shift in frequency.

\begin{figure}[h]
\begin{center}
\includegraphics[width=3.1in]{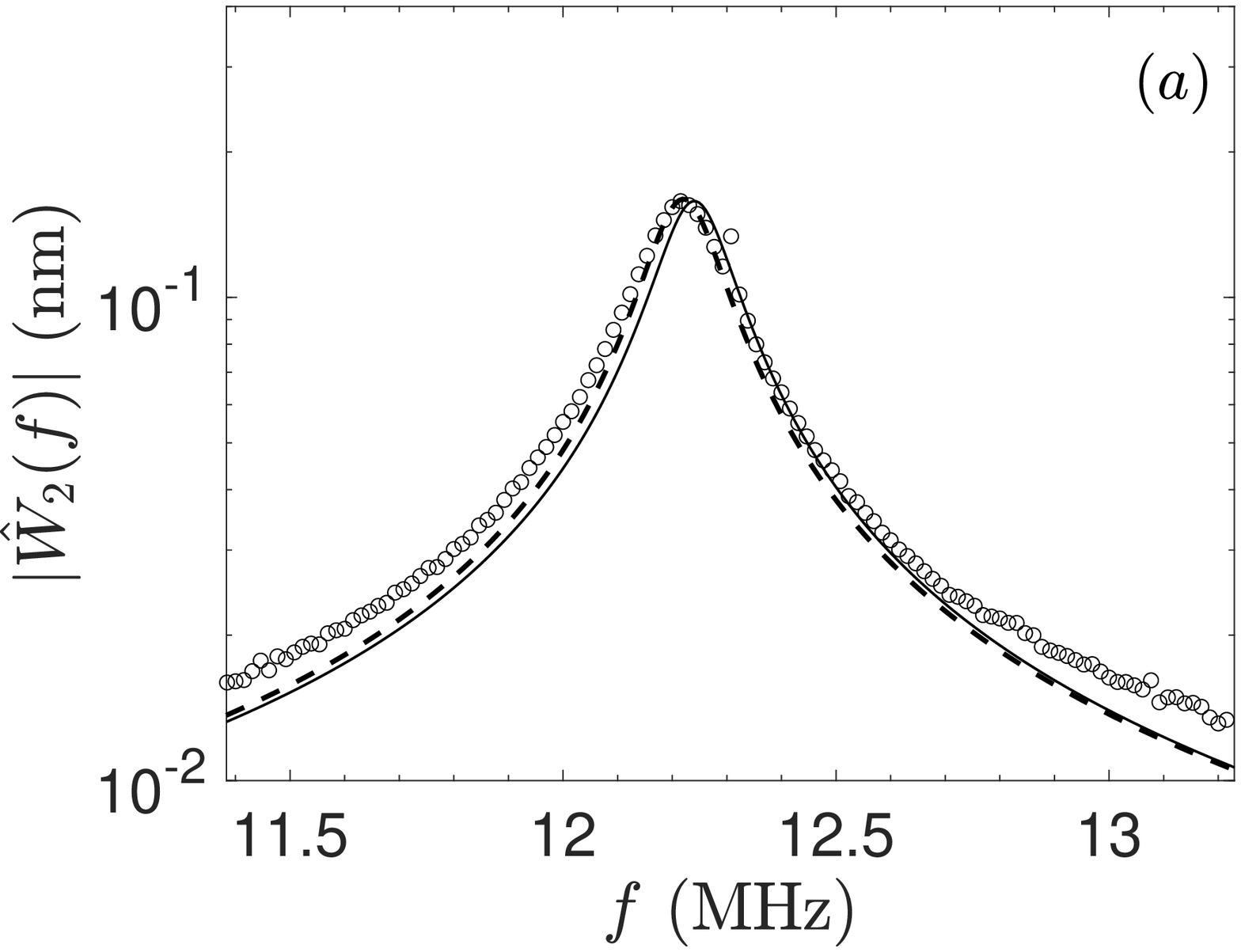}
\includegraphics[width=3.1in]{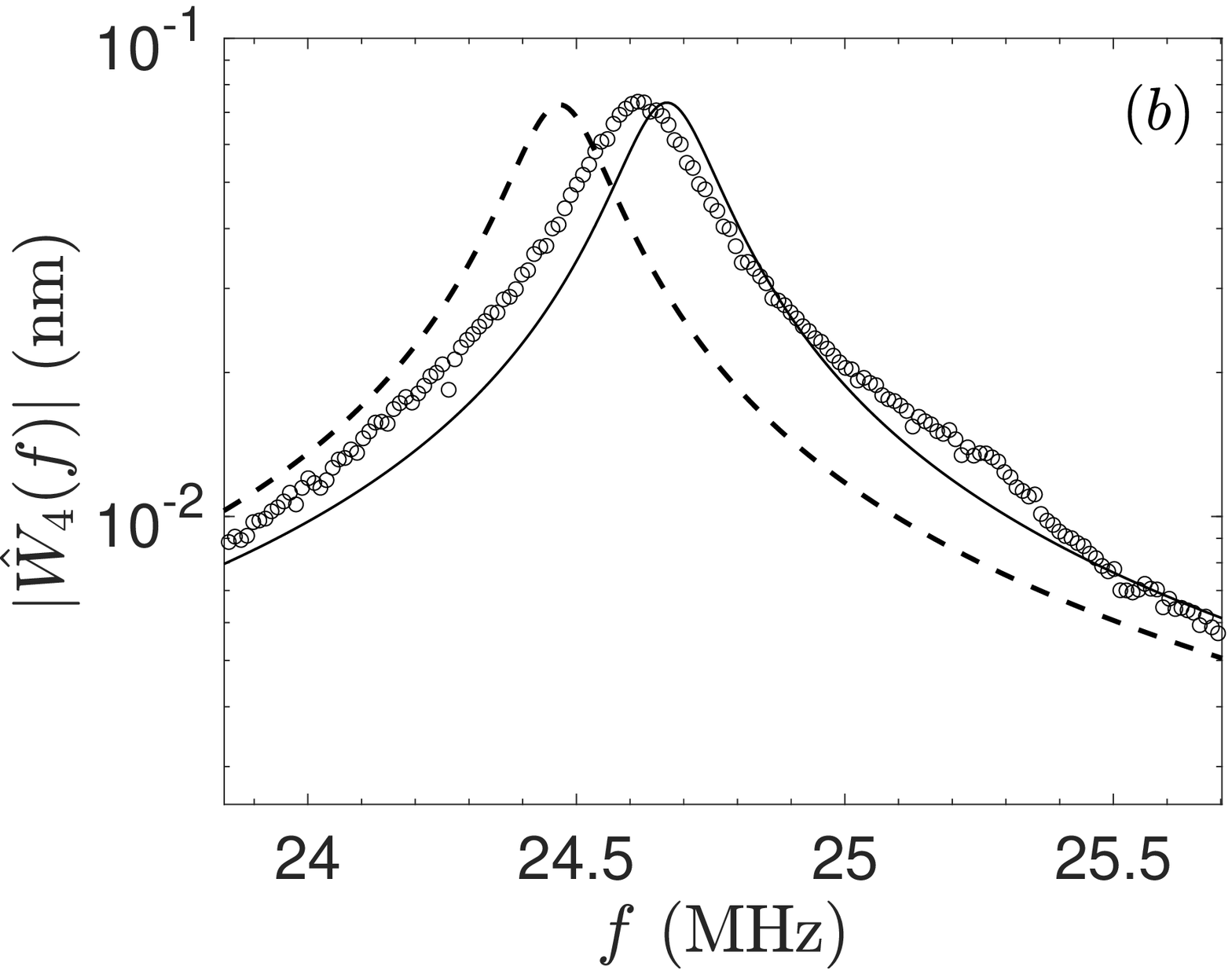}
\end{center}
\caption{The amplitude spectrum for the even modes when the beam is immersed in air. The open circles are the experimental measurements, the solid curve is the theoretical prediction using Eq.~(\ref{eq:whatnfinal-beam}) and $n\!=\!2$ and 4 and the dashed curve is the string prediction. (a) $n\!=\!2$, measured at $x_0\!=\!1/4$. (b) $n\!=\!4$, measured at $x_0\!=\!1/8$.}
\label{fig:theory-experiment-air-even}
\end{figure}

Figure~\ref{fig:theory-experiment-air-even} shows the amplitude spectra for the even modes where~(a) shows the second mode and (b)~shows the fourth mode. The agreement between the beam theory and experiment is excellent. The string description is also very good, however with the addition of a shift in frequency as expected.
%%%%%%%%%%%%%%%%%%%%%%%%%%%%%%%%%%%%
%%%%%%%%%%%%%%%%%%%%%%%%%%%%%%%%%%%%

\begin{figure}[h]
\begin{center}
\includegraphics[width=2.1in]{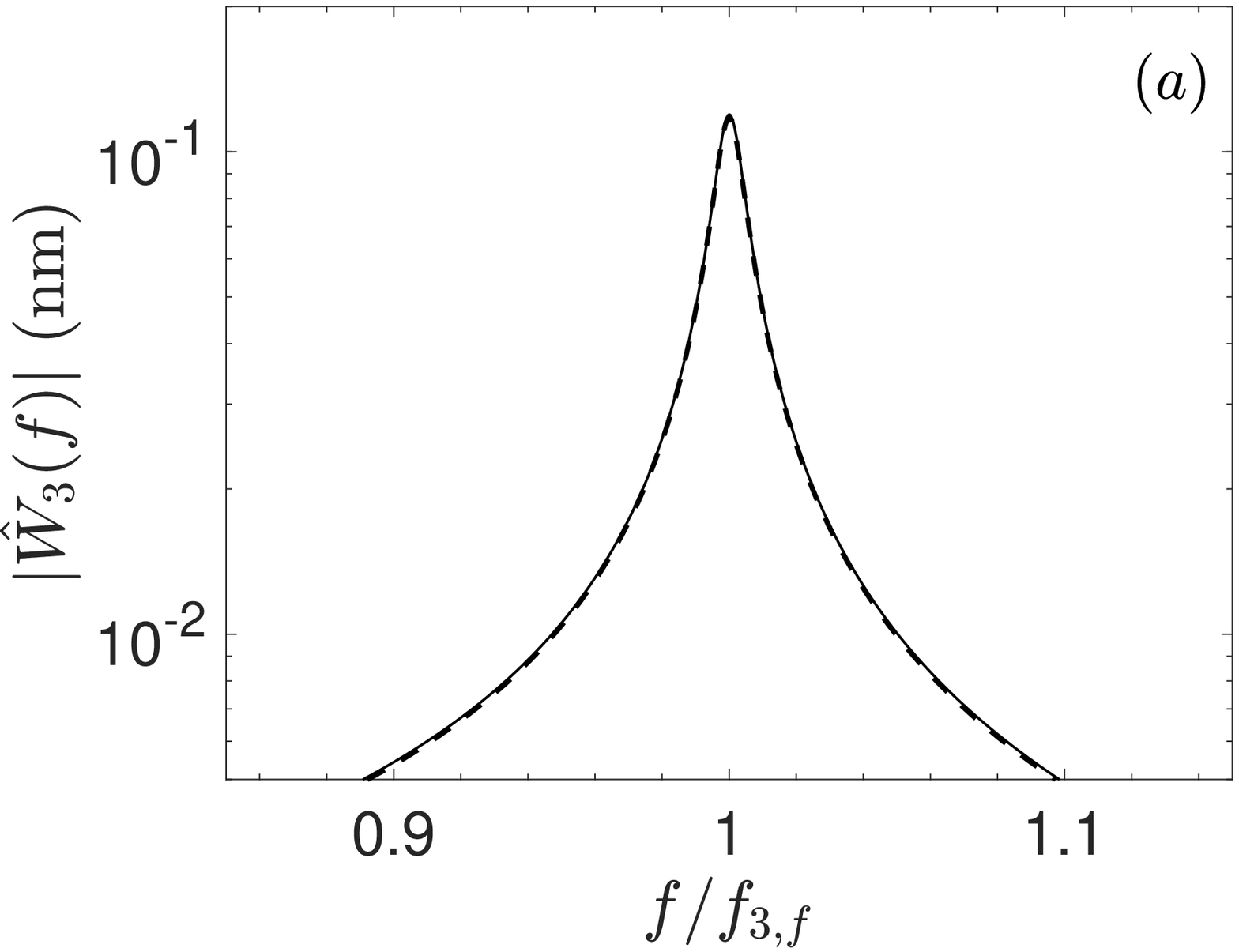}
\includegraphics[width=2.1in]{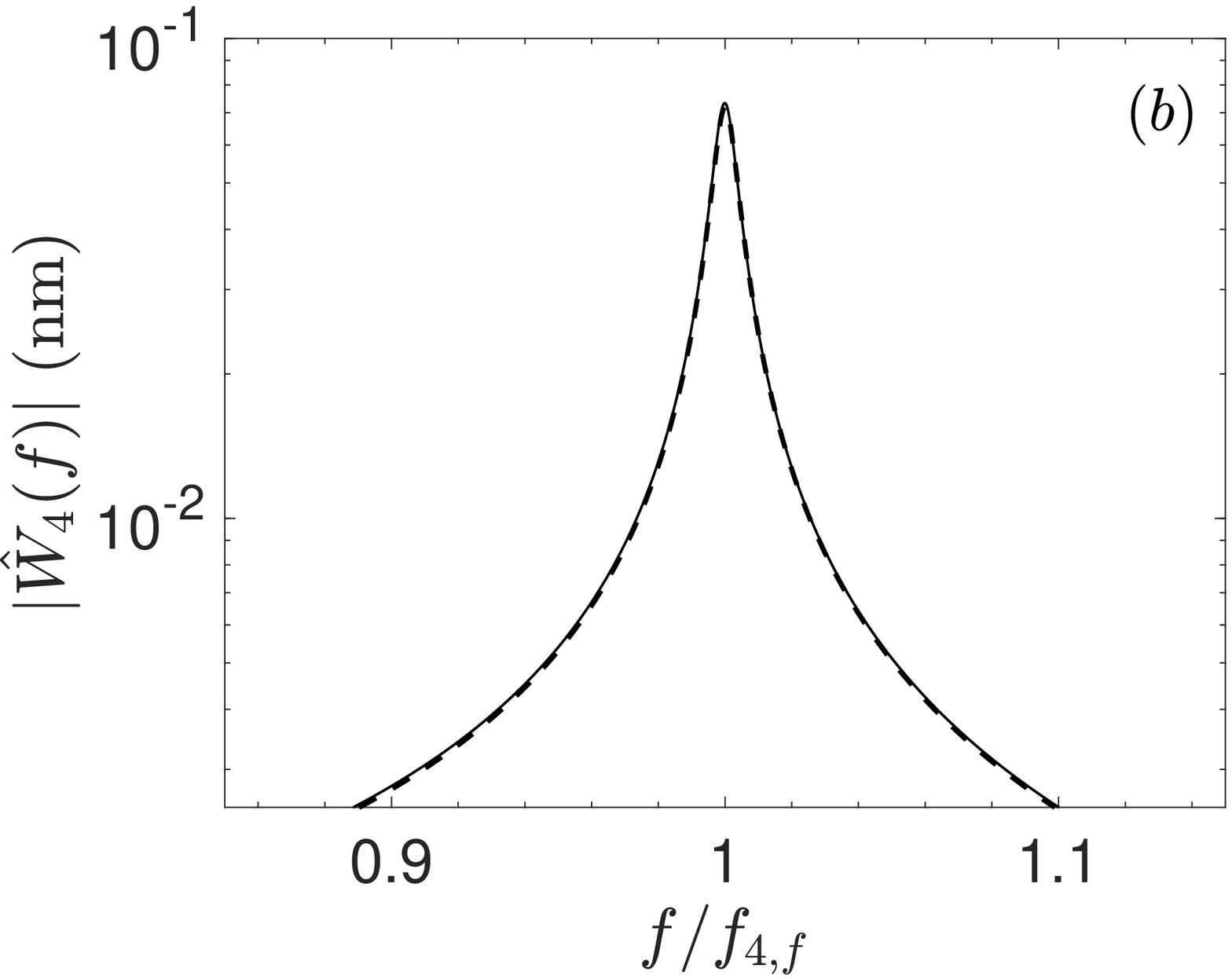}
\includegraphics[width=2.1in]{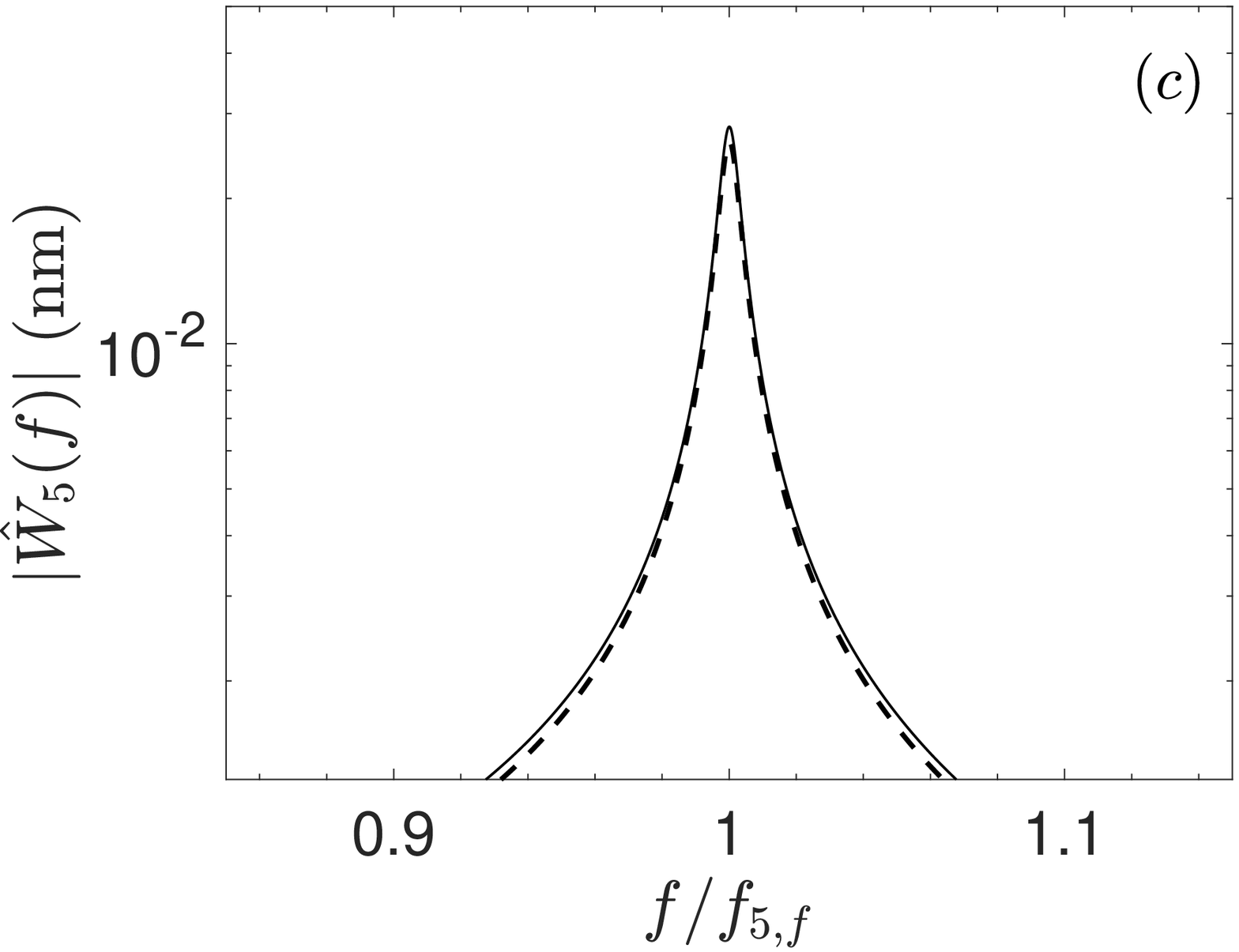}
\end{center}
\caption{A comparison of the beam and string theory descriptions of the amplitude spectra for the higher order modes of a beam immersed in air as a function of the normalized frequency. Solid lines are beam theory and dashed lines are the string prediction. The frequency, for each description, has been renormalized by its peak frequency in fluid $f_{n,f}$. (a) $n \!=\! 3$, (b) $n \!=\! 4$, (c) $n\!=\!5$.}
\label{fig:theory-experiment-air-normalized}
\end{figure}

The excellent agreement of the string description for the shape and magnitude of the amplitude spectra can be made more clear using a normalized frequency as shown in Fig.~\ref{fig:theory-experiment-air-normalized}.  The frequency has been normalized by the frequency of the amplitude peak of that mode in fluid such that the peak occurs at frequency of unity for each curve. The amplitude spectrum for the beam theory is the solid line and for the string theory it is the dashed line.  Figure~\ref{fig:theory-experiment-air-normalized} shows results for modes 3-5 in panels (a)-(c), respectively. The two descriptions are nearly indistinguishable when represented in this way. This clearly illustrates the usefulness of the string description for beams with very high tension.
%%%%%%%%%%%%%%%%%%%%%%%%%%%%%%%%%%%%%
%%%%%%%%%%%%%%%%%%%%%%%%%%%%%%%%%%%%%

\subsection{The beam immersed in water}
\begin{figure}[h]
\begin{center}
\includegraphics[width=3.1in]{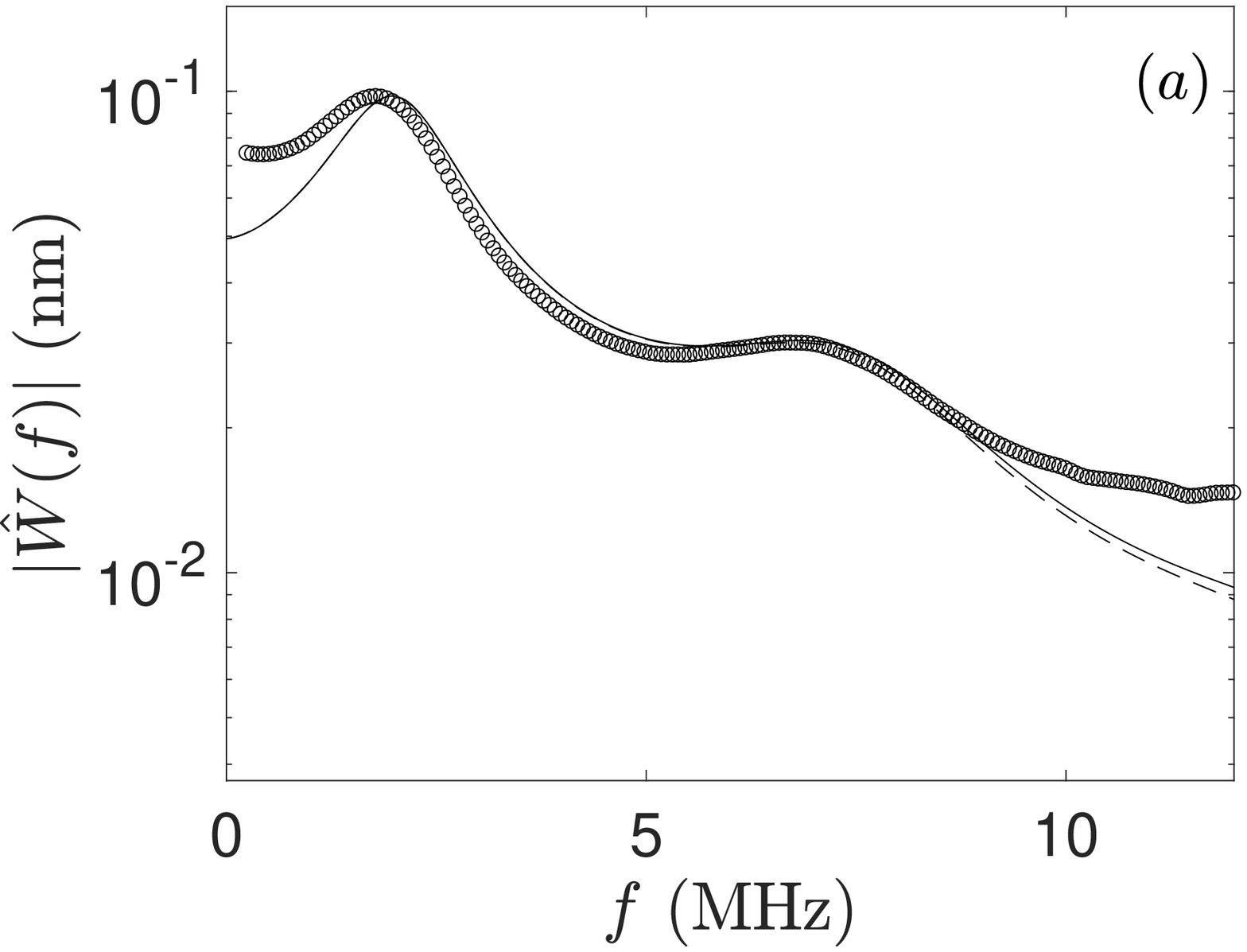}
\includegraphics[width=3.1in]{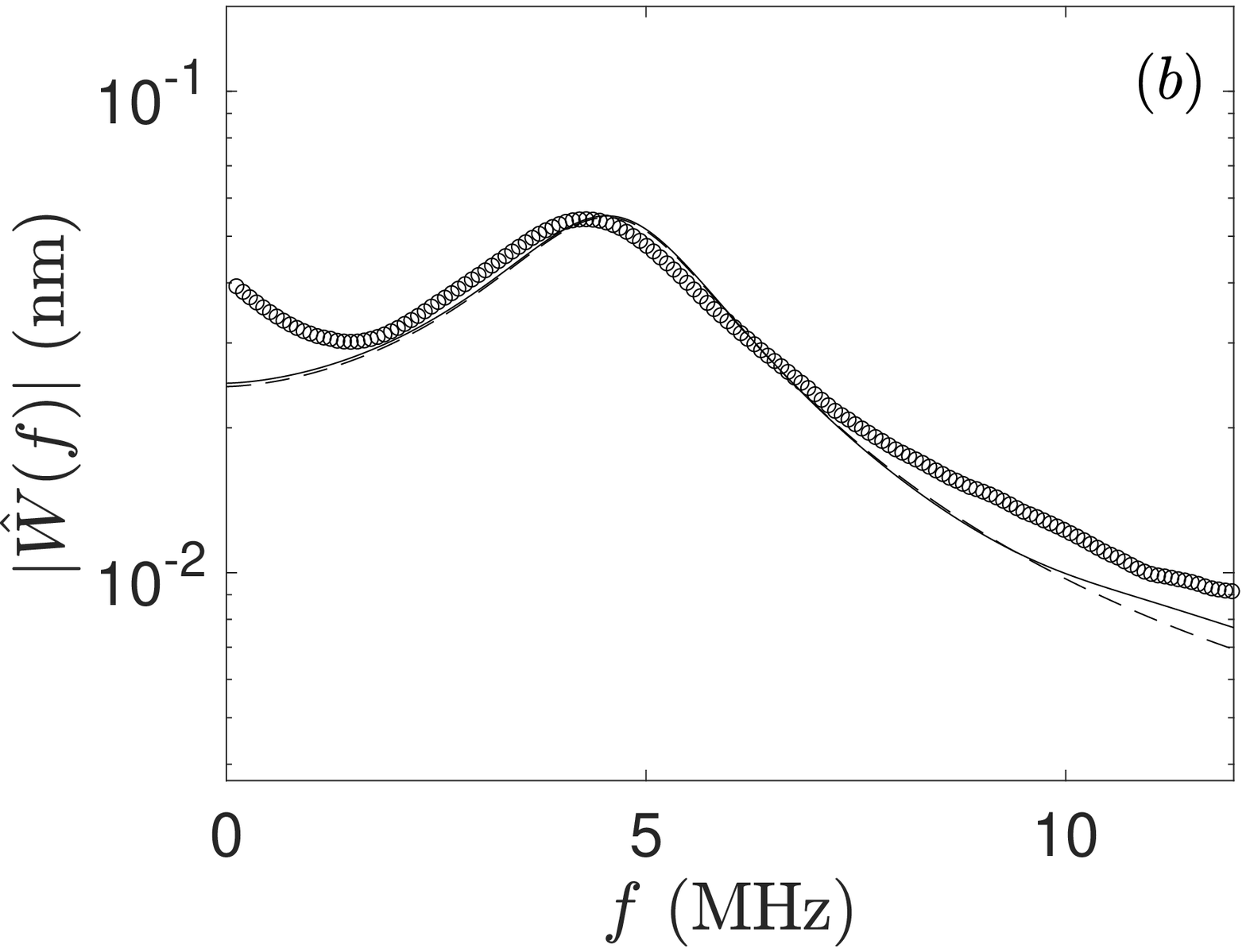}
\end{center}
\caption{The amplitude spectra for the beam when immersed in water. Open circles are experimental measurements, the solid line is the theoretical prediction using Eq.~(\ref{eq:whatfinal-beam}) from beam theory, and the dashed line is the string approximation using Eq.~(\ref{eq:whatfinal-string}). (a)~Modes 1 and 3 measured at $x_0\!=\!1/2$. In the theoretical prediction modes 1, 3, and 5 are included in the summation. (b)~Mode 2 measured at $x_0\!=\!1/4$. In the theoretical prediction modes 2, 4, and 6 are included in the summation.}
\label{fig:theory-experiment-water-odd-even}
\end{figure}

Figure~\ref{fig:theory-experiment-water-odd-even} shows the amplitude spectra of the beam when it is immersed in water. When the beam is immersed in water there is significant reduction in the frequencies of the peaks as well a broadening of the peaks. As a result of the peak broadening, the response from the different modes overlap significantly and it is not as useful to show the mode individually using Eq.~(\ref{eq:whatnfinal-beam}) but to include a representation of the full modal expansion given by Eq.~(\ref{eq:whatfinal-beam}).
%%%%%%%%%%%%%%%%%%%%%%%%%%%
%%%%%%%%%%%%%%%%%%%%%%%%%%%

Figure~\ref{fig:theory-experiment-water-odd-even}(a) shows the amplitude spectrum for the odd modes that are actuated by the symmetric drive and which are measured at $x_0\!=\!1/2$.  The experimental results are the open circles and the solid line is the theoretical prediction using Eq.~(\ref{eq:whatfinal-beam}) over odd modes where the infinite series is truncated at $n\!=\!5$.  The drastic reduction in the frequency of the fundamental mode is clearly evident with a value of 1.8 MHz. Mode 3 is also visible in this plot with a peak frequency of approximately 7 MHz. The string prediction is included as the dashed line which shows excellent agreement.  Since water has a much larger density and viscosity than that of air the spectra are dominated by the fluid properties. The small difference in the tension of the beam and string descriptions are less important as a result. The deviation between the theoretical predictions and the experimental measurement at very small frequency is a result of low frequency contributions in the experiments that are not included in our model.

Figure~\ref{fig:theory-experiment-water-odd-even}(b) shows the amplitude spectrum for mode 2 of the beam when driven asymmetrically and immersed in water. The experimental measurement of the beam deflection is at $x_0\!=\!1/4$. This location is near an anti-node for mode 4 and therefore it does not contribute here. The theoretical predictions use modal expansions that include modes 2, 4, and 6 in Eq.~(\ref{eq:whatfinal-beam}) for the beam with tension and in Eq.~(\ref{eq:whatfinal-string}) for the string. The agreement between the theory and experiment is very good.

\section{Conclusion}
\label{conclusion}

We have developed theoretical approaches to describe the dynamics of externally driven nanoscale beams that are under high tension and immersed in a viscous fluid. We have specifically focused upon doubly clamped beams that are driven externally using a spatially varying drive force applied near the ends of the beam. Our results are valid for the entire range of tension which includes the zero tension limit of the Euler-Bernoulli beam and the infinite tension limit of a string. By developing our approach as a modal expansion it is valid for the higher modes of the beam and it is also valid for measurements that are taken at any spatial location of the beam.  The string description reduces to much simpler expressions that we anticipate will be very useful in the design of future devices.

We have compared our theoretical predictions with an experiment that electrothermally drives the motion of a beam with high tension with a symmetric or asymmetric approach to drive the odd and even modes, respectively.  Our model yields quantitative agreement with the experiment for the first several modes of oscillation.  A more specialized model of the electrothermal driving would be needed for an improved description of this experiment and this would be an interesting avenue of future research.

It is important to highlight that full-scale numerical simulations of nanoscale beams immersed in a viscous fluid remain very computationally expensive. This is particularly true if complex driving mechanisms, such as an electrothermal drive, are included, and if multiple modes of oscillation are desired which require a large range of spatial and temporal scales to be resolved.  The theoretical description developed here provides important insights that will be useful for future experimental and theoretical efforts to explore these questions and nanoscale technologies further. 

\begin{acknowledgments}
\noindent M. R. Paul and J. Barbish acknowledge support from the National Science Foundation (NSF) Grant No. CMMI-2001559. K. L. Ekinci and C. Ti acknowledge support from the NSF Grant Nos. CMMI-1934271 and CMMI-2001403. All authors thank C. Yanik, I. I. Kaya and M. S. Hanay for fabricating the sample used in the measurements. 
\end{acknowledgments}

\noindent \textbf{Conflict of Interest}

\noindent The authors have no conflict to disclose.

\noindent \textbf{Data Availability}

\noindent The data that supports the findings of this study are available from the corresponding author upon reasonable request.

\end{document}